\documentclass[jimaging,review,accept,pdftex,moreauthors]{Definitions/mdpi} 
\firstpage{1} 
\pubvolume{9}
\issuenum{4}
\articlenumber{81}
\pubyear{2023}
\copyrightyear{2023}
\externaleditor{{Academic Editors: Cecilia Di Ruberto, Alessandro Stefano, Albert Comelli, Lorenzo Putzu and Andrea Loddo}}
\datereceived{13 March 2023} 
\daterevised{31 March 2023} 
\dateaccepted{7 April 2023} 
\datepublished{13 April 2023} 
\hreflink{https://doi.org/10.3390/\linebreak jimaging9040081} 

\makeatletter
\let\c@lofdepth\relax
\let\c@lotdepth\relax
\makeatother
\usepackage{subfigure}
\makeatletter
\renewcommand{\@thesubfigure}{(\textbf{\alph{subfigure}})~}
\makeatother

\Title{Deep Learning Approaches for Data Augmentation in Medical Imaging: A Review}

\TitleCitation{Deep Learning Approaches for Data Augmentation in Medical Imaging: A Review}

\Author{Aghiles Kebaili \orcidC{}, Jérôme Lapuyade-Lahorgue \orcidA{} and Su Ruan *\orcidB{}}

\AuthorNames{Aghiles Kebaili, Jerome Lapuyade-Lahorgue and Ruan Su}

\AuthorCitation{Kebaili 
, A.; Lapuyade-Lahorgue, J.; Ruan, S.}

\address[1]{Université Rouen Normandie, INSA Rouen Normandie, Université Le Havre Normandie, Normandie Univ, \mbox{LITIS UR 4108},  F-76000 Rouen, France
}

\corres{\hangafter=1 \hangindent=1.0em \hspace{-1em} {Correspondence: su.ruan@univ-rouen.fr}}

\abstract{Deep learning has become a popular tool for medical image analysis, but the limited availability of training data remains a major challenge, particularly in the medical field where data acquisition can be costly and subject to privacy regulations. Data augmentation techniques offer a solution by artificially increasing the number of training samples, but these techniques often produce limited and unconvincing results. To address this issue, a growing number of studies have proposed the use of deep generative models to generate more realistic and diverse data that conform to the true distribution of the data. In this review, we focus on three types of deep generative models for medical image augmentation: variational autoencoders, generative adversarial networks, and diffusion models. We provide an overview of the current state of the art in each of these models and discuss their potential for use in different downstream tasks in medical imaging, including classification, segmentation, and cross-modal translation. We also evaluate the strengths and limitations of each model and suggest directions for future research in this field. Our goal is to provide a comprehensive review about the use of deep generative models for medical image augmentation and to highlight the potential of these models for improving the performance of deep learning algorithms in medical image analysis.}


\begin{document}

\section{Introduction}

In recent years, advances in deep learning have been remarkable in many fields, including medical imaging. Deep learning is used to solve a wide variety of tasks such as classification \cite{amyar2022weakly,brochet2022quantitative}, segmentation \cite{zhou2022tri}, and detection \cite{chen2018unsupervised} using different types of medical imaging modalities, for instance, magnetic resonance imaging (MRI) \cite{lundervold2019overview}, computed tomography (CT) \cite{song2021deep}, and positron emission tomography (PET) \cite{islam2020gan}. Most of these modalities are defined as very high-dimensional data, and the number of training samples is often limited in the medical domain (e.g., the rarity of certain diseases). As deep learning algorithms rely on large amounts of data, running such applications in a low-sample-size regime can be very challenging. Data augmentation can increase the size of the training set by artificially synthesizing new samples. It is a very popular technique in computer vision~\cite{krizhevsky2017imagenet} and has become inseparable from deep learning applications when rich training sets are not available. Data generation is also used in the case of missing modalities for multimodal image segmentation \cite{zhou2022missing}. As a result, the model can be trained to generalize images with better quality and avoid overfitting. In addition, some deep learning frameworks, including PyTorch \cite{paszke2019pytorch}, allow for on-the-fly data augmentation during training, rather than physically expanding the training dataset. Basic data augmentation operations include random rotations, cropping, flipping, or noise injection. However, these simple operations are not sufficient when dealing with complex data such as medical images.



Several studies have been conducted to propose data augmentation schemes more suitable for the medical domain. The ultimate goal would be to reproduce a data distribution as close as possible to the real data, such that it is impossible, or at least difficult, to distinguish the newly sampled data from the real data. Recent performance improvements in deep generative models have made them particularly attractive for data augmentation. For example, generative adversarial networks (GANs) \cite{goodfellow2014generative} have demonstrated their ability to generate realistic images. As a result, this architecture has been widely used in the medical field \cite{sandfort2019data,mahapatra2019image} and has been included in several data augmentation reviews \cite{yi2019generative,ali2022role,chen2022generative}. Nevertheless, GANs also have their drawbacks, such as learning instability, difficulty in converging, and suffering from mode collapse \cite{mescheder2018training}, which is a state where the generator produces only a few samples. Variational autoencoders (VAEs) \cite{kingma2013auto} are another type of deep generative model that has received less attention in data augmentation. VAEs outperform GANs in terms of output diversity and are free of mode collapse. However, the major problem is their tendency to often produce blurry and hazy output images. This undesirable effect is due to the regularization term in the loss function. Recently, a new type of deep generative model called diffusion models (DMs) \cite{sohl2015deep,ho2020denoising} has emerged and promises remarkable results with a great ability to generate realistic and diverse outputs. However, DMs are still in their infancy and are not yet well established in the medical field, but are expected to be a promising alternative to previous generative models. One of the drawbacks of DMs is their high computational cost and huge sampling time.

\begin{figure}[H]
    \begin{adjustwidth}{-\extralength}{0cm}
    \subfigure[]{\includegraphics[width=7.8cm]{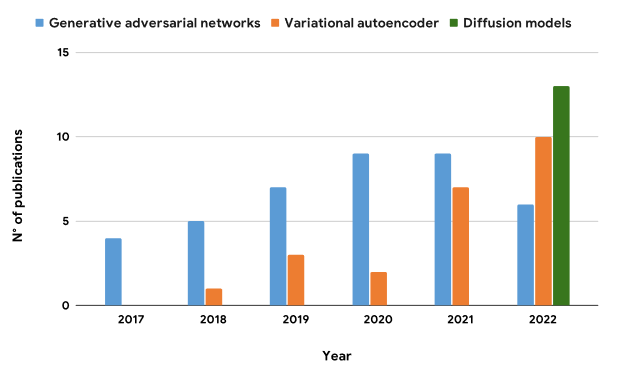}}
    \subfigure[]{\includegraphics[width=7.8cm]{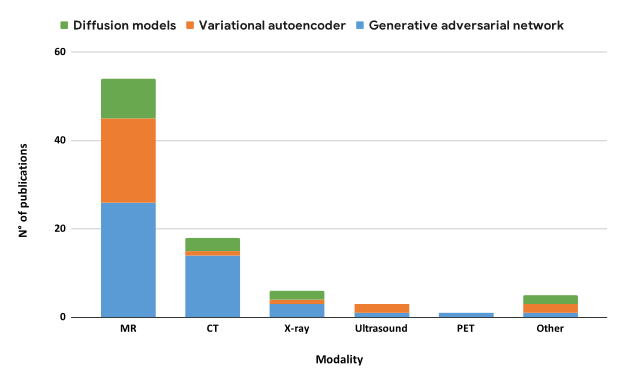}}
    \linebreak \centering
    \subfigure[]{\includegraphics[width=6.8cm]{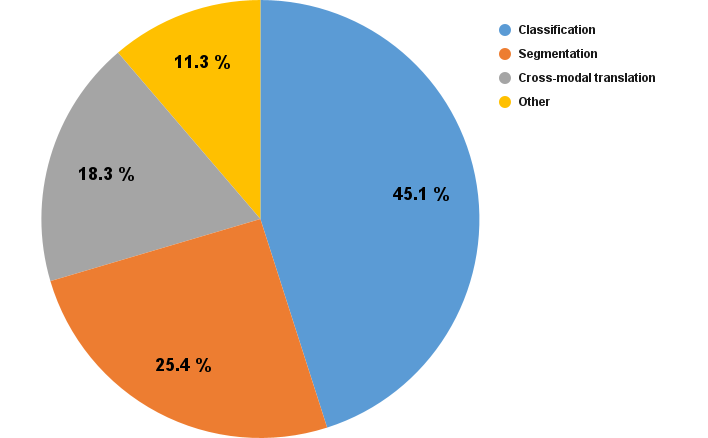}}
    \end{adjustwidth}
    \caption{Distribution 
    of publications on deep generative models applied to medical imaging data augmentation as of 2022. (\textbf{a}) The number of publications per architecture type and year. (\textbf{b}) The distribution of publications by modality, with CT and MRI being the most-commonly studied imaging modalities. Note that for cross-modal translation tasks, both the source and target modalities are counted in this plot. (\textbf{c}) The distribution of publications by downstream task, with segmentation and classification being the most common tasks in medical imaging. This figure illustrates the increasing interest in using deep generative models for data augmentation in medical imaging and highlights the diversity of tasks and modalities that have been addressed in the literature.\label{fig:stats}}
\end{figure}

Different approaches have been proposed to solve this generative learning trilemma of quality sampling, fast sampling, and diversity \cite{xiao2021tackling}. In this paper, we review the state of the art of deep learning architectures for data augmentation, focusing on three types of deep generative models for medical image augmentation: VAEs, GANs, and DMs. To provide an accurate review, we harvested a large number of publications via the PubMed and Google Scholar search engines. We selected only publications dating from at least 2017 using various keywords related to data augmentation in medical imaging. Following this, a second manual filtering was performed to eliminate all cases of false positives (publications not related to the medical field and/or data augmentation). In conclusion, 72 publications have been kept, mainly from journals such as IEEE Transactions In Medical Imaging or Medical Image Analysis and conferences such as Medical Image Computing and Computer Assisted Intervention and IEEE International Symposium on Biomedical Imaging. Some publications will be described in more detail in Section \ref{sec3}; these have been selected according to two criteria: date of publication and number of citations. Nevertheless, all the articles are available in descriptive tables as well as other information such as datasets used to perform training. These different papers were organized according to the deep generative model employed and the main downstream tasks targeted by the generated data (i.e., classification, segmentation, and cross-modal translation). Knowing the dominance of GANs for data augmentation in the medical imaging domain, the objective of this article is to highlight other generative models. To the best of our knowledge, this is the first review article that compares different deep generative models for data augmentation in medical imaging and does not focus exclusively on GANs \cite{yi2019generative,ali2022role,chen2022generative}, nor traditional data augmentation methods~\cite{chlap2021review,shorten2019survey}. The quantitative ratio of GAN-based articles to the rest of the deep generative models is very unbalanced; nevertheless, we try to bring some equilibrium to this ratio in the hope that an unbiased comparative study following this paper may be possible in the future. To further illustrate our findings, we present a graphical representation of the selected publications in Figure \ref{fig:stats}. This figure provides a comprehensive overview of the number of publications per year, per modality, and per downstream task. By analyzing these graphics, we can observe the trends and the preferences of the scientific community in terms of the use of deep generative models for data augmentation in medical imaging.

This article is organized as follows: Section \ref{sec2} presents a brief theoretical view of the above deep generative models. Section \ref{sec3} reviews deep generative models for medical imaging data augmentation, grouped by the targeted application. Section \ref{sec4} discusses the advantages and disadvantages of each architecture and proposes a direction for future research. Finally, Section \ref{sec5} concludes the paper.

\section{Background}\label{sec2}
\begin{figure}[H]
    \begin{adjustwidth}{-\extralength}{0cm}
    \subfigure[]{\includegraphics[width=7.8cm]{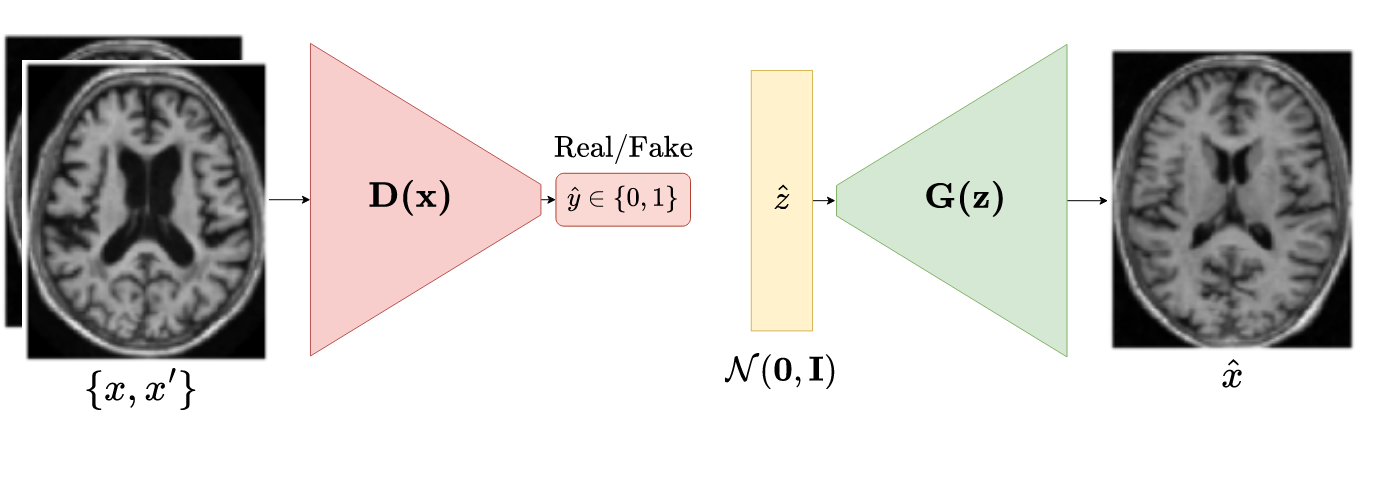}}
    \subfigure[]{\includegraphics[width=7.8cm]{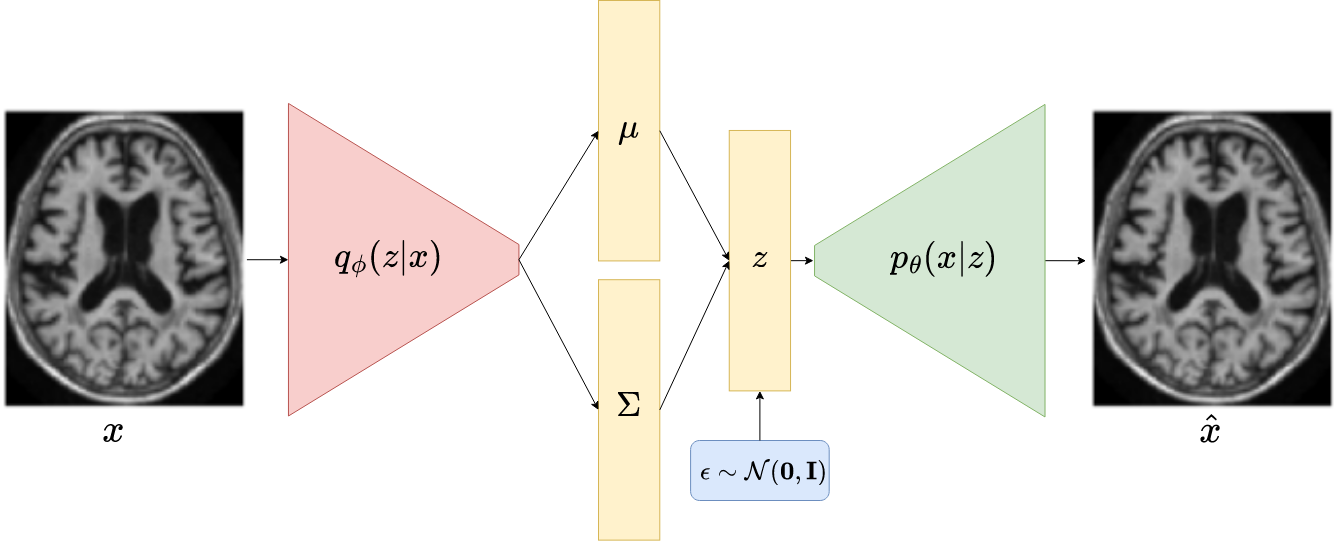}}
    \linebreak \centering
    \subfigure[]{\includegraphics[width=11.8cm]{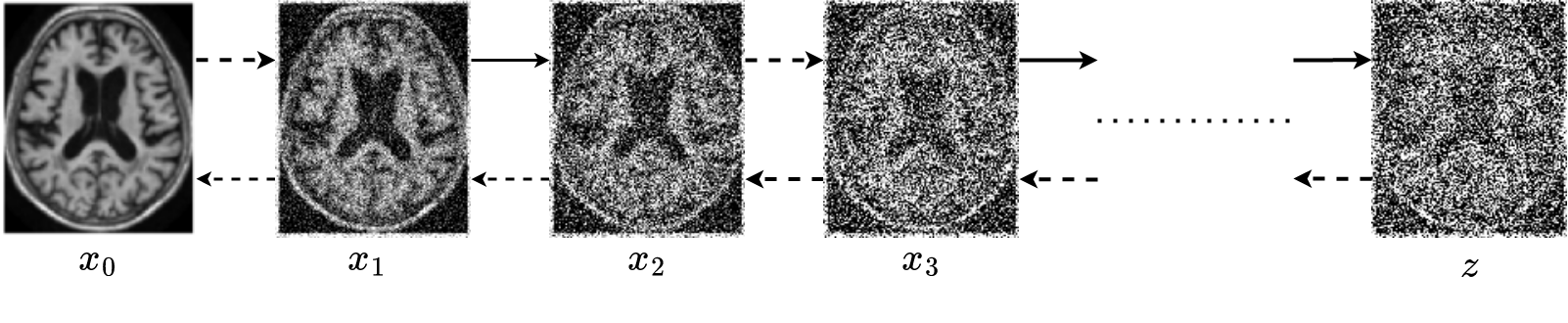}}
    \end{adjustwidth}
    \caption{Illustration 
 of the three deep generative models that are commonly used for medical image augmentation: (\textbf{a}) generative adversarial networks (GANs), which consist of a generator and a discriminator network trained adversarially to generate realistic data; (\textbf{b}) variational autoencoders (VAEs), which consist of an encoder and a decoder network trained to reconstruct data and learn a compact latent representation; and (\textbf{c}) diffusion models, which consist of a forward and backward flow of information through a series of steps to model the data distribution.} \label{fig:deep generative models}
\end{figure}

The main goal of deep generative models is to learn the underlying distribution of the data and to generate new samples that are similar to the real data. Our deep generative model can be represented as a function $g : z \longrightarrow x$ that maps a low-dimensional latent vector $z \in \mathbb{R}^d$ to a high-dimensional data point $x \in \mathbb{R}^D$ such as $d \leq D$. The latent variable $z$ is a realization of a random vector that is sampled from a prior distribution $p(z)$. The data point $x$ is another realization sampled from the data distribution $p(x)$. The goal of the deep generative model is to learn the mapping function $g$ such that the generated data $g(z)$ are similar to the real data $x$ associated with $z$. Each deep generative model proposes its own approach to learn the mapping function $g$. In this section we present a brief overview of the most popular deep generative models. Figure \ref{fig:deep generative models} provides a visual representation of their respective architectures.

\subsection{Generative Adversarial Networks}
GAN \cite{goodfellow2014generative} is a class of deep generative models composed of two separate networks: a generator and a discriminator. The generator can be seen as a mapping function $G$ from a random latent vector $z$ to a data point $x$, where $z$ is sampled from a fixed prior distribution $p(z)$ commonly modelled as a Gaussian distribution. The discriminator $D$ is a binary classifier that takes a data point $x$ as input and outputs a probability $D(x)$ such that $x$ is a real data point. During the training process, the generator $G$ is trained to replicate data points $x_g$ so that the discriminator cannot distinguish between real data points $x_r$ and the generated data points $x_g$. On the other hand, the discriminator $D$ is trained to differentiate the fake from the real data points. Those two networks are trained simultaneously in an adversarial manner, hence the name generative adversarial network. The loss functions of $G$ and $D$ can be expressed as follow : 

\begin{equation}
\begin{aligned}
    \mathcal{L}_G &= \min_{\theta} \mathbb{E}_{z \sim p(z)}[\log D_\phi(G_\theta(z))] \\
    \mathcal{L}_D &= \max_{\phi} \mathbb{E}_{x \sim p(x)}[\log D_\phi(x)] + \mathbb{E}_{z \sim p(z)}[\log (1 - D_\phi(G_\theta(z)))]
\end{aligned}
\end{equation}
where $\theta$ and $\phi$ are the corresponding learnable parameters for the generator and discriminator neural networks, respectively.

This adversarial learning has proven to be effective in capturing the underlying distribution of the real data distribution $p(x)$. This has been inspired by game theory and can be seen as a minimax game between the generator and the discriminator. It is ultimately desirable to reach a Nash equilibrium where both the generator and discriminator are equally effective at their tasks. The loss function can be summarized as follows:

\begin{equation}
\mathcal{L}_{GAN} = \min_{\theta} \max_{\phi} \mathbb{E}_{x \sim p(x)}[\log D_\phi(x)]  + \mathbb{E}_{z \sim p(z)}[\log (1 - D_\phi(G_\theta(z)))]
\end{equation}

Once trained, new data points can be synthesized by sampling a random latent vector $z$ from the prior distribution $p(z)$ and feeding it to the generator.

\subsection{Variational Autoencoders}
Variational inference is a Bayesian inference technique that allows us to estimate the posterior distribution $p(z|x)$ with a simpler distribution $q(z|x)$.
The aim of variational inference is to minimize a Kullback--Leibler divergence between the posterior distribution $p_\theta(z|x)$ and the variational distribution $q_\phi(z|x)$, where $\theta$ and $\phi$ are the posterior and variational distribution parameters, respectively. The Kullback--Leibler is the most commonly used. The loss function based on Kullback--Leibler is defined as follows:

\begin{equation} 
\min_{\theta,\phi} D_{KL}(q_\phi(z|x) || p_\theta(z|x)) = \min_{\theta,\phi} \mathbb{E}_{z \sim q_\phi}[\log \frac{q_\phi(z|x)}{p_\theta(z|x)}]
\end{equation} 

With further simplifications, and applying Jensen's inequality, we can rewrite the above equation as:
\vspace{-6pt}
\begin{equation}
    \log p_\theta(x) = -\mathbb{E}_{z\sim q_\phi}[\log q_\phi(z|x)] + \mathbb{E}_{z \sim q_\phi}[\log p_\theta(z, x)]
    + D_{KL}(q_\phi(z|x) || p_\theta(z|x))
\end{equation}

\begin{equation} 
\begin{aligned}
    \log p_\theta(x) &\geq  -\mathbb{E}_{z\sim q_\phi}[\log q_\phi(z|x)] + \mathbb{E}_{z\sim q_\phi}[\log p_\theta(z, x)] \\
    &\geq \mathbb{E}_{z\sim q_\phi}[\log p_\theta(x|z)] - \mathbb{E}_{z\sim q_\phi}[\log \frac{q_\phi(z|x)}{p(z)}] = ELBO
\end{aligned}
\end{equation}
where $\log p_\theta(x)$ is the marginal log likelihood of the data $x$, $p(z)$ is the prior distribution of the latent variable $z$, generally modeled as a Gaussian distribution, and ELBO is the evidence lower bound. The variational distribution $q_\phi(z|x)$ can be learned by minimizing $D_{KL}(q_\phi(z|x) || p_\theta(z|x))$, which is equivalent to maximizing the ELBO given a fixed $\theta$. This ELBO term can be further decomposed into two terms: the reconstruction term and the regularization term. The reconstruction term measures the difference between the input data and its reconstruction, and it is typically calculated using binary cross-entropy loss. The regularization term ensures that the latent variables follow a desired distribution, such as a normal distribution, and it is calculated using the Kullback--Leibler divergence between the latent distribution and the desired distribution. Together, these two terms form the ELBO loss function, which is used to train the VAE model. The VAE is composed of an encoder $q_\phi(z|x)$ and a decoder $p_\theta(x|z)$. The encoder $q_\phi(z|x)$ is a neural network that maps the data $x$ to the latent variable $z$. The decoder $p_\theta(x|z)$ is a neural network that maps the latent variable $z$ to the data $x$. The VAE is trained by minimizing the reconstruction and regularization terms \eqref{lrec}.
\vspace{-6pt}
\begin{equation}
    \mathcal{L}_{rec} = \mathbb{E}_{q_\phi(z|x)}[\log p_\theta(x|z)], \;\;
    \mathcal{L}_{reg} = \mathbb{E}_{q_\phi(z|x)}[\log \frac{q_\phi(z|x)}{p(z)}]
\label{lrec}
\end{equation}

Once trained, new data points can be synthesized by sampling a random latent vector $z$ from the prior distribution $q_\phi$ and feeding it to the decoder. In other words, the decoder represents the generative model.

\subsection{Diffusion Probabilistic Models}
Diffusion models \cite{sohl2015deep,ho2020denoising} are a class of generative models that are based on the diffusion process. The diffusion process is a stochastic process that can be seen as a parameterized Markov chain. Each transition in the chain gradually adds a Gaussian noise to an initial data point $x_0$ of distribution $q(x)$. The diffusion process can be expressed as follow:
\begin{equation}
\begin{aligned}
    q(x_t|x_{t-1}) &= \mathcal{N}(\sqrt{\alpha_t}x_{t-1}, \beta_t \text{I})\\
    q(x_{1:T}|x_0) &= \prod_{t=1}^T q(x_t|x_{t-1})
\end{aligned}
\end{equation}
where $\beta_t \in [0, 1], t = 1, \ldots, T$ is the predefined noise variance at step $t$, $\alpha_t = 1 - \beta_t$, and $T$, the total number of steps. The diffusion model is trained to reverse the diffusion process starting with a noise input $x_T \sim \text{N}(\text{0}, \text{I})$ and reconstructing the initial data point $x_0$. This denoising process can be seen as a generative model. The reverse diffusion process can be expressed as follows:
\vspace{-6pt}
\begin{equation}
p_\theta(x_{0:T}) = p(x_T)\prod_{t=1}^{T} p_\theta(x_{t-1}|x_t), \;\; q(x_{t - 1}|x_{t}) = \mathcal{N}(\mu(x_t, t), \Sigma(x_t, t))
\end{equation}
where $\mu(x_t, t)$ and $\Sigma(x_t, t)$ are the mean and the variance of the denoising model at step $t$. Similarly to the VAE, diffusion models learn to recreate the true sample at each step by maximizing the evidence lower bound (ELBO), matching the true denoising distribution $q(x_{t-1}|x_t)$ and the learned denoising distribution $p_\theta(x_{t-1}|x_t)$. By the end of the training, the diffusion model will be able to map a noise input $x_T$ to the initial data point $x_0$ throught reverse diffusion; hence, new data points can be synthesized by sampling a random noise vector $x_T$ from the prior distribution $\mathcal{N}(0, \text{I})$ and feeding it to the model.

\subsection{Exploring the Trade-Offs in Deep Generative Models: The Generative Learning Trilemma}
\subsubsection{Generative Adversarial Networks}
The design and training of VAEs, GANs, and DMs is often subject to trade-offs between fast sampling, high-quality samples, and mode coverage, known as the generative learning trilemma \cite{xiao2021tackling}. Among these models, GANs have received particular attention due to their ability to generate realistic images and are the first deep generative models to be extensively used for medical image augmentation. They are known for their ability to generate high-quality samples that are difficult to distinguish from real data. However, they may suffer from mode collapse, a phenomenon where the model only generates samples from a limited number of modes or patterns in the data distribution, potentially leading to poor coverage of the data distribution and a lack of diversity in the generated samples. To address mode collapse, several variations of  GAN have been proposed. One popular approach is the Wasserstein GAN (WGAN) \cite{arjovsky2017wasserstein}, which replaces the Jensen--Shannon divergence used in the original GAN with the Wasserstein distance, a metric that measures the distance between two probability distributions. This has the benefit of improving the quality of the generated samples. Another widely used extension is the conditional GAN (CGAN) \cite{mirza2014conditional}, which adds a conditioning variable $y$ to the latent vector $z$ in the generator, allowing for more control over the generated samples and partially mitigating mode collapse. The CGAN can be seen as a generative model that can generate data points $x$ conditioned on $y$ and models the joint distribution $p(x, y)$. A GAN with a conditional generator has been introduced by Isola et al. \cite{pix2pix} to learn to translate images from one domain to another by replacing the traditional noise-to-image generator with a U-Net \cite{ronneberger2015u}. The adversarial learning process allows the U-Net to generate more realistic images based on a better understanding of the underlying data distribution.

\begin{figure}[H]
\includegraphics[width=11cm]{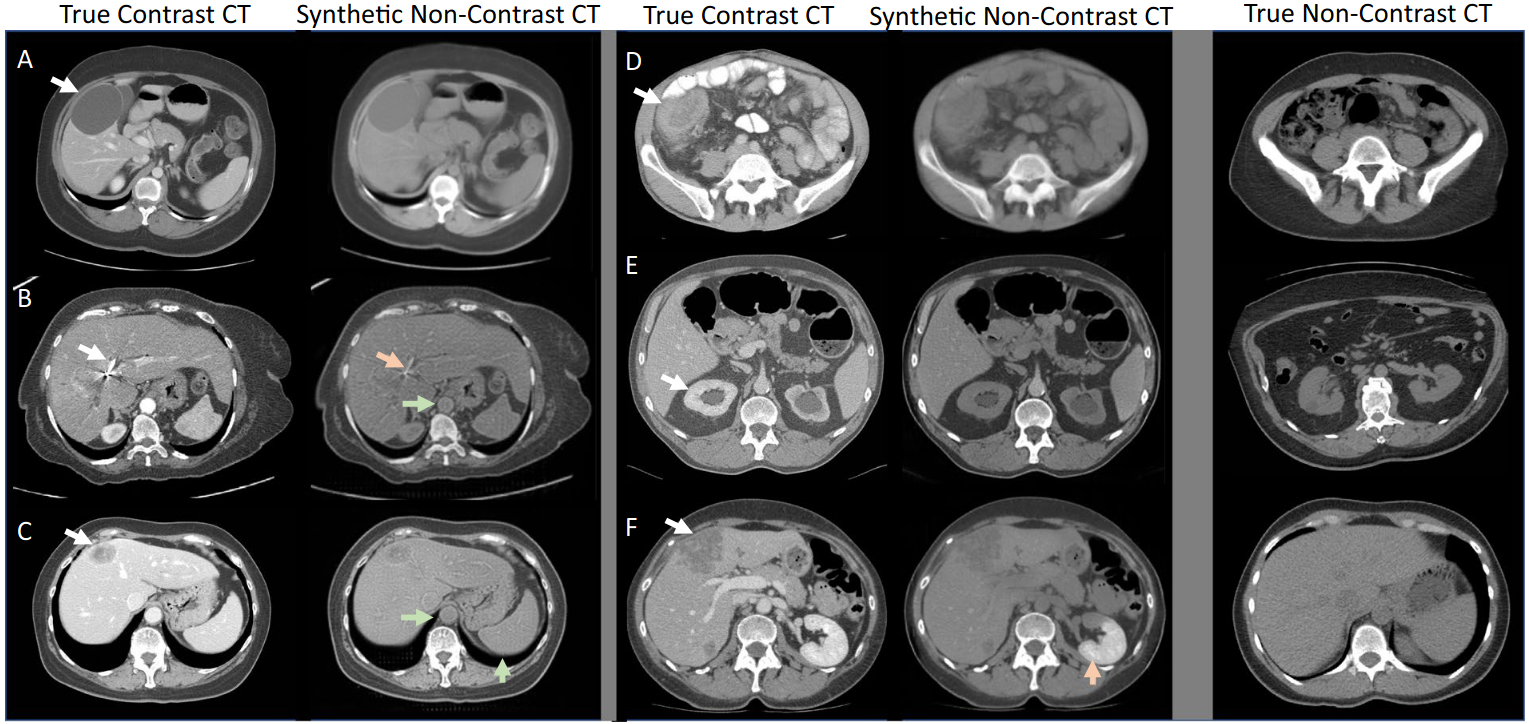}
\caption{Adapted 
 from Sandfort et al. \cite{sandfort2019data}, the study presented examples of true contrast CT scans and synthetic non-contrast CT scans generated using a CycleGAN. \textls[-20]{The left columns show the true contrast CT scans, while the right columns present the synthetic non-contrast CT scans. It is observed that the synthetic non-contrast images generated with CycleGAN appeared convincing, even in the presence of significant abnormalities in the contrast CT scans. The last column on the right displays unrelated examples of non-contrast images. The letters A to F in this figure represent various abnormalities/pathologies, and the arrows indicate their corresponding synthetic non-contrast CT images. However, they are not essential for understanding the main purpose of the figure, which is to demonstrate the generator's ability to produce realistic images.}\label{fig:gan}}
\end{figure}

\textls[-15]{Other variations of the GAN include deep convolutional GAN (DCGAN) \cite{dcgan}, progressive growing GAN (PGGAN) \cite{karras2017progressive}, CycleGAN \cite{cyclegan}, auxiliary classifier GAN (ACGAN)~ \cite{odena2016conditional}, VAE-GAN \cite{vaegan}, and many others, which have been proposed to address various issues such as training stability, scalability, and quality of the generated samples. While these variants have achieved good results in a variety of tasks, they also come with their own set of trade-offs. Despite these limitations, GANs are generally fast at generating new images, making them a good choice for data augmentation when well-trained. As an example, Figure \ref{fig:gan} showcases the capacity of a CycleGAN to generate realistic synthetic medical images.}

\subsubsection{Variational Autoencoders}
VAEs are a another type of deep generative model that has gained popularity for their ease of training and good coverage of the data distribution. Unlike GANs, VAEs are trained to maximize the likelihood of the data rather than adversarially, making them a good choice for tasks that require fast sampling and good coverage of the data distribution. Using variational inference methods, VAEs are able to better approximate the real data distribution given a random noise vector, thus making them less vulnerable to mode collapse. Moreover, VAEs enable the extraction of relevant features and can learn a smooth latent representation of the data, which allows for the interpolation of points in the space providing more control over the generated samples \cite{higgins2016early}. 

VAEs have not been as commonly used for data augmentation compared to GANs due to the blurry and hazy nature of the generated samples. However, several proposals, such as inverse autoregressive flow \cite{iafvae}, InfoVAE \cite{zhao2017infovae}, or VQ-VAE2 \cite{vqvae2}, have been made to improve the quality of VAE-generated samples as well as the variational aspect of the model. Despite this, most of these extensions have not yet been applied to medical image augmentation. A more effective approach to addressing the limitations of VAEs in this context is to utilize a hybrid model called a VAE-GAN, which combines the strengths of both VAEs and GANs to generate high-quality, diverse, and realistic synthetic samples. While VAE-GANs cannot fully fix the low-quality generation of VAEs, they do partially address this issue by incorporating the adversarial training objective of GANs, which allows for the improvement of visual quality and sharpness of the generated samples while still preserving the ability of VAEs to learn a compact latent representation of the data. In addition to VAE-GANs, another common architecture for medical image augmentation is the use of conditional VAEs (CVAEs), which allows for the control of the output samples by conditioning the generation process on additional information, such as class labels or attributes. This can be particularly useful in medical imaging, as it allows for the generation of synthetic samples that are representative of specific subgroups or conditions within the data. By using conditional VAEs, it is possible to generate synthetic samples that are more targeted and relevant to specific tasks or analyses. In summary, VAEs, VAE-GANs, and conditional VAEs are all viable approaches for medical image augmentation, each offering different benefits and trade-offs in terms of diversity, quality, and fidelity of the generated~samples.

\subsubsection{Diffusion Models}
There has been a recent surge in the use of DMs for image synthesis in the academic literature due to their superior performance in generating high-quality and realistic synthesized images compared to other deep generative models such as VAEs and GANs \cite{dhariwal2021diffusion}. This success can be attributed to the way in which DMs model the data distribution by approximating it using a series of simple distributions combined through the diffusion process, allowing them to capture complex, high-dimensional distributions and generate samples that are highly representative of the underlying data. This is especially useful for synthesizing images as natural images often have a wide range of textures, colors, and other visual features that can be difficult to model using simpler parametric models. This can also be applied to medical imaging where data tends to be complex. However, DMs can also have some limitations, such as being computationally intensive to solve, especially for large or complex systems, and requiring a significant amount of data to be accurately calibrated. In addition, DMs have a long sampling time compared to other deep generative models such as VAEs and GANs due to the high number of steps in the reverse diffusion process (ranging from several hundreds to thousands). This issue is compounded when the model is being used in real-time applications or when it is necessary to generate large numbers of samples. As a result, researchers have proposed several solutions and variants of diffusion models that aim to improve the sampling speed while maintaining high-quality and diverse samples. {\color{black} These include 
strategies such as progressive distillation \cite{salimans2022progressive}. This method involves distilling a trained deterministic diffusion sampler, using many steps, into a new diffusion model that takes half as many sampling steps. Another way to improve the sampling time is the use of improved variants such as Fast Diffusion Probabilistic Model (FastDPM) \cite{kong2021fast}, which uses a modified optimization algorithm to reduce the sampling time and introduces a concept of continuous diffusion process, or with non-Markovian diffusion models such as Denoising Diffusion Implicit Model (DDIM) \cite{song2020denoising}}. Similarly to VAE-GAN, ref. \cite{xiao2021tackling} proposes the denoising diffusion GAN, which is a hybrid architecture between DMs and multimodal conditional GANs \cite{mirza2014conditional}, which have been shown to produce high-quality and diverse samples at a much faster sampling speed compared to the original diffusion models (factor of $\times$2000). Overall, while diffusion models have demonstrated great potential in the field of image synthesis, their long sampling time remains a challenge that researchers are actively working to address.

\section{Deep Generative Models for Medical Image Augmentation}\label{sec3}

Medical image processing and analysis using deep learning has developed rapidly in the past years, and it has been able to achieve state-of-the-art results in many tasks. However, the lack of data is still a major issue in this field. To address this, medical image augmentation became a crucial task, and many studies have been conducted in this direction. In this section, we will review the different deep generative models that have been proposed to generate synthetic medical images. This review is organized into three different categories corresponding to each one of the  deep generative models. The publications are further classified according to the downstream task targeted by the generated images. We address here the most common tasks in medical imaging: classification, segmentation, and cross-model image translation, which will be summarized in the form of tables.

\subsection{Generative Adversarial Networks}
As part of their study, Han et al. \cite{han2018gan} proposed the use of two variants of GANs for generating (2D) MRI sequences: a WGAN \cite{arjovsky2017wasserstein} and a DCGAN \cite{dcgan}, in which combinations of convolutions and batch normalizations replace the fully-connected layers. The results of this study were presented in the form of a visual Turing test where an expert physician was asked to classify real and synthetic images. For all MRI sequences except FLAIR images, WGAN was significantly more successful at deceiving the physician than DCGAN (62\% compared to 54\%). The same author further proposes using PGGAN \cite{karras2017progressive} combined with traditional data augmentation techniques such as geometric transformations. PGGAN is a GAN with a multi-stage training strategy that progressively increases the resolution of the generated images. The results indicate that combining PGGAN with traditionally augmented data can slightly improve the performance of the classifier when compared to using  PGGAN alone.

\textls[-20]{Conditional synthesis is a technique that allows the generation of images conditioned on a specific variable $y$. This is particularly useful in medical imaging, where tasks such as segmentation or cross-modal translation are widespread. A variable $y$ serves as the ground truth for the generated images and can be expressed in various ways, including class labels, segmentation maps, or translation maps. In this context, Frid-Adar et al. \cite{frid2018gan} propose to use an ACGAN \cite{odena2016conditional} for synthesizing liver lesions in CT images. The ACGAN is a GAN with a discriminator conditioned on a class label. Three label classes were considered: cysts, metastases, and hemangiomas. Based solely on conventional data augmentation, the classification results produced a sensitivity of 78.6\% and a specificity of 88.4\%. By adding the synthetic data augmentation, the results increased to a sensitivity of 85.7\% and a specificity of 92.4\%. \mbox{Guibas et al. \cite{guibas2017synthetic}} propose a two-stage pipeline for generating synthetic images of fundus photographs with associated blood vessel segmentation masks. In the first stage, synthetic segmentation masks are generated using DCGAN, and in the second stage, these synthetic masks are translated into photorealistic fundus images using CGAN. Comparing the Kullback--Leibler divergence between the real and synthetic images revealed no significant differences between the two distributions. In addition, the authors evaluated the generated images on a segmentation task using only synthetic images, showing an F1 score of 0.887 versus 0.898 when using real images. This negligible difference indicates the quality of the generated images. By the same token, Platscher et al. \cite{platscher2020image} propose using a two-step image translation approach to generate MRI images with ischemic stroke lesion masks. The first step consists of generating synthetic stroke lesion masks using a WGAN. The newly generated fake lesions are implanted on healthy brain anatomical segmentation masks. Finally, those segmentation masks are fed into a pretrained image-translation model that maps the mask into a real ischemic stroke MRI. The authors studied three different image translation models, CycleGAN \cite{cyclegan}, Pix2Pix~\cite{pix2pix}, and SPADE \cite{park2019semantic}, and reported that Pix2Pix was the most successful in terms of visual quality. A U-Net \cite{ronneberger2015u} was trained using both clinical and generated images and showed an improvement in the Dice score compared to the model trained only on clinical images (63.7\% to 72.8\%).}

Regarding cross-modal translation, Yurt et al. \cite{yurt2021mustgan} propose a multi-stream approach for generating missing or corrupted MRI contrasts from other high-quality ones using a GAN-based architecture. The generator is composed of multiple one-to-one streams and a joint many-to-one stream, which are designed to learn latent representations sensitive to unique and common features of the source, respectively. The complementary feature maps generated in the one-to-one streams and the shared feature maps generated in the many-to-one stream are combined with a fusion block and fed to a joint network that infers the final image. In their experiments, the authors compare their approach to other state-of-the-art translation GANs and show that the proposed method is more effective in terms of quantitative and radiological assessments. The synthesized images presented in this study demonstrate the effectiveness of deep learning approaches applied to data augmentation in medical imaging. Specifically, the study investigated two tasks: (a) T1-weighted image synthesis from T2- and PD-weighted images and (b) PD-weighted image synthesis from T1- and T2-weighted images. The results obtained from the proposed method outperformed other variants of GANs such as pGAN \cite{dar2019image} and MM-GAN \cite{sun2020mm}, highlighting its effectiveness for image synthesis in medical imaging.

In summary, the use of GANs for data augmentation has been demonstrated to be a successful approach. The studies discussed in this section have employed some of the most innovative and known GAN architectures in the medical field, including WGAN, DCGAN, and Pix2Pix, and have primarily focused on three tasks: classification, segmentation, and cross-modal translation. Custom-made GAN variants have also been proposed in the current state of the art (see Table \ref{tab:gan}), some of which could be explored further. Notably, conditional synthesis has proven to be particularly useful for tasks such as segmentation and cross-modal translation, as seen with the ACGAN and Pix2Pix, resulting in an improved classification performance. Additionally, two-stage pipeline approaches have been proposed for generating synthetic images conditioned on segmentation masks. To further illustrate the use of GANs for medical image augmentation, we present a summary of the relevant studies in Table \ref{tab:gan}. This table includes information about the dataset, imaging modality, and evaluation metrics used in each study, as well as the specific type of GAN architecture employed. A further discussion will be presented in Section \ref{sec4}.

\begin{table}[H]
\tablesize{\footnotesize}
\caption{Overview of GAN-based architectures for medical image augmentation, including hybrid status of architectures (if applicable), indicating used combinations of VAEs, GANs, and DMs.\label{tab:gan}}

\setlength{\cellWidtha}{\fulllength/7-2\tabcolsep-0.2in}
\setlength{\cellWidthb}{\fulllength/7-2\tabcolsep+0.2in}
\setlength{\cellWidthc}{\fulllength/7-2\tabcolsep-0in}
\setlength{\cellWidthd}{\fulllength/7-2\tabcolsep+0.1in}
\setlength{\cellWidthe}{\fulllength/7-2\tabcolsep-0in}
\setlength{\cellWidthf}{\fulllength/7-2\tabcolsep-0.6in}
\setlength{\cellWidthg}{\fulllength/7-2\tabcolsep+0.5in}
	   \begin{adjustwidth}{-\extralength}{0cm}
		\begin{tabularx}{\fulllength}{>{\raggedright\arraybackslash}m{\cellWidtha}>{\raggedright\arraybackslash}m{\cellWidthb}>{\raggedright\arraybackslash}m{\cellWidthc}>{\raggedright\arraybackslash}m{\cellWidthd}>{\raggedright\arraybackslash}m{\cellWidthe}>{\raggedright\arraybackslash}m{\cellWidthf}>{\raggedright\arraybackslash}m{\cellWidthg}}
        
		\toprule 
            \textbf{Reference}	& 
            \textbf{Architecture} &
            \textbf{Hybrid Status} & 
            \textbf{Dataset} &
            \textbf{Modality} &
            \textbf{3D} &
            \textbf{Eval. Metrics}\\
		\midrule 
            \textbf{Classification} &&&&&&\\
            \midrule
            \cite{frid2018gan} & DCGAN, ACGAN & & Private & CT & & Sens., Spec.\\
            \cite{han2018gan} & DCGAN, WGAN & & BraTS2016 & MR & & Acc.\\
            \cite{han2019combining} & PGGAN, MUNIT & & BraTS2016 & MR & \checkmark & Acc., Sens., Spec.,\\
            \cite{kwon2019generation} & AE-GAN & Hybrid (V~+~G)& BraTS2018, ADNI & MR & \checkmark & MMD, MS-SSIM\\
            \cite{zhuang2019fmri} & ICW-GAN & & OpenfMRI, HCP & MR & \checkmark & Acc., Prec., F1\\  
            &&& NeuroSpin, IBC &&&Recall\\ 
            \cite{waheed2020covidgan} & ACGAN & & IEEE CCX & X-ray & &Acc., Sens., Spec.\\
            &&&&&&Prec., Recall, F1 \\ 
            \cite{han2020infinite} & PGGAN & & BraTS2016 & MR & & Acc., Sens., Spec.\\
            \cite{sun2020adversarial}  & ANT-GAN & & BraTS2018 & MR & & Acc. \\
            \cite{wang2020class}  & MG-CGAN & & LIDC-IDRI & CT & & Acc., F1 \\
            \cite{geng2020deep} & FC-GAN & Hybrid (V~+~G) & ADHD, ABIDE & MR & &Acc., Sens., Spec., AUC\\
            \cite{pang2021semi} & TGAN & & Private & Ultrasound & & Acc., Sens., Spec.\\
            \cite{barile2021data} & AAE & & Private & MR & & Prec., Recall, F1\\
            \cite{shen2021mass} & DCGAN, InfillingGAN & & DDSM & CT & & LPIPS, Recall\\
            \cite{ambita2021covit} & SAGAN & & COVID-CT, SARS-COV2 & CT & & Acc.\\
            \cite{hirte2021realistic} & StyleGAN & & Private & MR & & -\\
            \cite{kaur2021mr}  & DCGAN & & PPMI & MR & & Acc., Spec., Sens.\\
            \cite{guan2022medical} & TMP-GAN & & CBIS-DDMS, Private & CT & & Prec., Recall, F1, AUC\\
            \cite{ahmad2022brain} & VAE-GAN & Hybrid (V~+~G) & Private & MR & & Acc., Sens., Spec.\\
            \cite{pombo2022equitable} & CounterSynth & & UK Biobank, OASIS & MR & \checkmark & Acc., MSE, SSIM, MAE\\
            \midrule
            \textbf{Segmentation} &&&&&&\\
            \midrule
            \cite{guibas2017synthetic} & CGAN & & DRIVE & Fundus photography & & KLD, F1\\
            \cite{neff2017generative} & DCGAN & & SCR & X-ray & & Dice, Hausdorff\\
            \cite{mok2018learning} & CB-GAN & & BraTS2015 & MR & & Dice, Prec., Sens.\\
            \cite{shin2018medical} & Pix2Pix & & BraTS2015, ADNI & MR &\checkmark& Dice\\
            \cite{sandfort2019data} & CycleGAN & & NIHPCT & CT & & Dice\\
            \cite{jiang2019cross} & CM-GAN & & Private & MR & & KLD, Dice \\
            &&&&&&hausdorff \\ 
            \cite{jiang2020covid} & CGAN & & COVID-CT &CT & & FID, PSNR, SSIM, RMSE\\
            \cite{qasim2020red} & Red-GAN & & BraTS2015, ISIC & MR & & Dice\\
            \cite{platscher2020image} & Pix2Pix, SPADE, CycleGAN & & Private & MR & & Dice\\
            \cite{shi2020novel} & StyleGAN & & LIDC-IDRI & CT & & Dice, Pres., Sens.\\
            \cite{shen2022image} & DCGAN, GatedConv & & Private & X-ray & & MAE, PSNR, SSIM, FID, AUC\\

            \midrule
            
            \textbf{Cross-modal translation} &&&&&&\\
            \midrule
            \cite{chartsias2017adversarial} & CycleGAN & & Private & MR $\leftrightarrow$ CT & \checkmark & Dice\\
            \cite{wolterink2017deep} & CycleGAN & & Private & MR $\rightarrow$ CT & & MAE, PSNR\\
            \cite{nie2018medical} & Pix2Pix & & ADNI, Private & MR $\rightarrow$ CT & \checkmark & MAE, PSNR, Dice\\
            \cite{armanious2020medgan}  & MedGAN & & Private & PET $\rightarrow$ CT & & SSIM, PSNR, MSE\\
            &&&&&& VIF, UQI, LPIPS \\ 
            \cite{dar2019image}  & pGAN, CGAN & & BraTS2015, MIDAS, IXI & T1 $\longleftrightarrow$ T2 & & SSIM, PSNR\\
            \cite{jiang2019cross}  & CM-GAN & & Private & MR & & KLD, Dice \\
            &&&&&&hausdorff \\ 
            \cite{yurt2021mustgan}  & mustGAN & & IXI, ISLES & T1 $\leftrightarrow$ T2 $\leftrightarrow$ PD & & SSIM, PSNR\\
            \cite{yang2021synthesizing} & CAE-ACGAN & Hybrid (V~+~G) & Private & CT $\rightarrow$ MR & \checkmark & PSNR, SSIM, MAE\\
            \cite{sikka2021mri} & GLA-GAN & & ADNI & MR $\rightarrow$ PET & & SSIM, PSNR, MAE\\
            &&&&&& Acc., F1 \\ 
            \midrule
            \textbf{Other} & & & \\
            \midrule
            \cite{amirrajab2022pathology}  & VAE-CGAN & Hybrid (V~+~G) & ACDC & MR & \checkmark & -\\
		\bottomrule
		\end{tabularx}
	\end{adjustwidth}
	\noindent{\footnotesize{Note: V = variational autoencoders, G = generative adversarial networks.}}
\end{table}

\subsection{Variational Autoencoders}
Zhuang et al. \cite{zhuang2019fmri} present an empirical evaluation of 3D functional MRI data augmentation using deep generative models such as VAEs and GANs. The results indicate that CVAE and conditional WGAN can produce diverse, high-quality brain images. A 3D convolutional neural network (CNN) was used to further evaluate the generated samples on the original and augmented data in a classification task, demonstrating an accuracy improvement of $3.17\%$ when using CVAE augmented data and $3.72\%$ when using CWGAN augmented data. As part of Pesteie et al. \cite{pesteie2019adaptive}, a revised variant of the CVAE is proposed, called the ICVAE, which separates the embedding space of the input data and the conditioning variables. This allows the generated image characteristics to be independent of the conditioning variables, resulting in a more diverse output. In contrast, the standard CVAE encodes the data and conditioning variables in a shared embedding space. The authors evaluate the ICVAE on classification and segmentation tasks using transverse ultrasound images of the spine and FLAIR MRI images of the brain, respectively. The results demonstrate an improvement of $8.0\pm1.0\%$ in classification accuracy and $4.5\pm0.5\%$ in the Dice score compared to the model trained on real images only. The ICVAE model is able to generate more realistic MRI images by encoding appearance features independently of the structures in its latent space. The authors demonstrate the generation of synthetic MRI and ultrasound images using the ICVAE architecture, which are conditioned on a tumor segmentation mask and a label indicating the center-line of the spine, respectively. The CVAE architecture is also shown for comparison. Chadebec et al. \cite{chadebec2022data} introduce a novel Geometry-aware VAE for high dimensional data augmentation in low sample size settings. This model combines Riemannian metric learning with normalizing flows to improve the expressiveness of the posterior distribution and learn meaningful latent representations of the data. Additionally, the authors propose a new non-prior sampling scheme based on Hamiltonian Monte Carlo, since the standard procedure utilizing the prior distribution is highly dependent upon the data, especially for small datasets. As a result, the generated samples are remarkably more realistic than those generated by a conventional VAE, and the model is more resilient to the lack of data. An evaluation of the synthetic data on a classification task shows an improvement in accuracy from 66.3\% to 74.3\% using 50 real + 5000 synthetic MRIs, compared to using only the original data. The original paper by Chadebec et al. \cite{chadebec2022data} includes a challenge in which readers are invited to identify the real brain MRIs from fake ones.

Other studies suggest the use of VAEs to improve the segmentation task performance. Huo et al. \cite{huo2022brain} introduce a progressive VAE-based architecture (PAVAE) for generating synthetic brain lesions with associated segmentation masks. The authors propose a two-step pipeline where the first step consists in generating synthetic segmentation masks based on a conditional adversarial VAE. The CVAE is assisted by a ``condition embedding block'' that encodes high-level semantic information of the lesion into the feature space. The second step involves generating photorealistic lesion images conditioned on the lesion mask using ``mask embedding blocks'', which encodes the lesion mask into the feature space during generation, similar to SPADE. The authors compare their approach to other state-of-the-art methods and show that PAVAE can produce more realistic synthetic lesions with associated segmentation masks. A segmentation network is trained using both real and synthetic lesions and shows an improvement in the Dice score compared to the model trained only on real images (66.69\% to 74.18\%).

In a recent paper, Yang et al. \cite{yang2021synthesizing} propose a new model for cross-domain translation called conditional variational autoencoding GAN (CAE-ACGAN).  CAE-ACGAN combines the advantages of both VAEs and GANs in a single end-to-end architecture. The integration of VAE and GAN, along with the implementation of an auxiliary discriminative classifier network, allows for a partial resolution of the challenges posed by image blurriness and mode collapse. Moreover, the VAE incorporates skip connections between the encoder and decoder, which enhances the quality of the images generated. In addition to translating 3D CT images into their  corresponding MR, the CAE-ACGAN generates more realistic images as a result of its discriminator, which serves as a quality-assurance mechanism. Based on PSNR and SSIM scores, the CAE-ACGAN model showed a mild improvement over other state-of-the-art architectures, such as Pix2Pix and WGAN-GP \cite{gulrajani2017improved}.

Table \ref{tab:vae} compiles a summary of the relevant studies using VAEs in medical data augmentation. In contrast to GANs, the number of studies employing VAEs for data augmentation in medical imaging is relatively low. However, almost half of these studies have utilized hybrid architectures, combining VAEs with adversarial learning. Interestingly, we observe that unlike GANs, there are not many VAE variants in medical imaging. Most commonly used VAE architectures are either conditional, such as vanilla CVAE and ICVAE, or hybrid architectures, such as IntroVAE, PAVAE, and ALVAE. Further discussion on the effectiveness of VAEs for medical image augmentation and the specific architectures utilized in previous studies will be presented in Section \ref{sec4}.

\begin{table}[H]
\tablesize{\small}
\caption{\textls[-15]{Overview of VAE-based architectures for medical image augmentation, including hybrid status of architectures (if applicable), indicating the combination of VAEs and GANs used in each study}.\label{tab:vae}} 
\setlength{\cellWidtha}{\fulllength/7-2\tabcolsep-0in}
\setlength{\cellWidthb}{\fulllength/7-2\tabcolsep+0.1in}
\setlength{\cellWidthc}{\fulllength/7-2\tabcolsep-0in}
\setlength{\cellWidthd}{\fulllength/7-2\tabcolsep-0in}
\setlength{\cellWidthe}{\fulllength/7-2\tabcolsep-0in}
\setlength{\cellWidthf}{\fulllength/7-2\tabcolsep-0.5in}
\setlength{\cellWidthg}{\fulllength/7-2\tabcolsep+0.4in}
	   \begin{adjustwidth}{-\extralength}{0cm}
		\begin{tabularx}{\fulllength}{>{\raggedright\arraybackslash}m{\cellWidtha}>{\raggedright\arraybackslash}m{\cellWidthb}>{\raggedright\arraybackslash}m{\cellWidthc}>{\raggedright\arraybackslash}m{\cellWidthd}>{\raggedright\arraybackslash}m{\cellWidthe}>{\raggedright\arraybackslash}m{\cellWidthf}>{\raggedright\arraybackslash}m{\cellWidthg}}
        
		\toprule 
%
            \textbf{Reference}	& 
            \textbf{Architecture} &
            \textbf{Hybrid Status} & 
            \textbf{Dataset} &
            \textbf{Modality} &
            \textbf{3D} &
            \textbf{Eval. Metrics}\\
		\midrule 
            \textbf{Classification} & & & \\
            \midrule
            \cite{pesteie2019adaptive} & ICVAE & & Private & MR & & Acc., Sens., Spec.\\
            &&&& Ultrasound &&Dice, Hausdroff, $\ldots$\\
            \cite{zhuang2019fmri}  & CVAE & & OpenfMRI, HCP & MR & \checkmark & Acc., Prec., F1\\  
            &&& NeuroSpin, IBC &&&Recall\\ 
            \cite{chadebec2022data}	& GA-VAE & & ADNI, AIBL & MR & \checkmark & Acc., Spec., Sens.\\
            \cite{terzopoulos2019multi}	& MAVENs & Hybrid (V~+~G) & APCXR & X-ray & & FID, F1\\
            \cite{hirte2021realistic} & IntroVAE & Hybrid (V~+~G)& Private & MR & & -\\
            \cite{qiang2021modeling}	& DR-VAE & & HCP & MR & & -\\
            \cite{ahmad2022brain}  & VAE-GAN & Hybrid (V~+~G) & Private & MR & & Acc., Sens., Spec.\\
            \cite{madan2022synthetic} & VAE & & Private & MR & & Acc.\\
            \cite{chadebec2021data} & RH-VAE & & OASIS &MR & \checkmark & Acc.\\

            \midrule
            \textbf{Segmentation} & & & \\
            \midrule
            \cite{liang2021data} 	& VAE-GAN & Hybrid (V~+~G) & Private & Ultrasound & & MMD, 1-NN, MS-SSIM\\
            \cite{gan2022esophageal}& AL-VAE & Hybrid (V~+~G) & Private & OCT \textsuperscript{1} & & MMD, MS, WD\\
            \cite{huo2022brain} 	& PA-VAE & Hybrid (V~+~G) & Private & MR & \checkmark & PSNR, SSIM, Dice \\
            &&&&&& NMSE, Jacc., $\ldots$\\
            \midrule
            
            \textbf{Cross-modal translation} & & & \\
            \midrule
            \cite{yang2021synthesizing}  & CAE-ACGAN & Hybrid (V~+~G) & Private & CT $\rightarrow$ MR & \checkmark & PSNR, SSIM, MAE\\
            \cite{hu2022domain}	& 3D-UDA & & Private & FLAIR $\leftrightarrow$ T1 $\leftrightarrow$ T2 & \checkmark & SSIM, PSNR, Dice\\

            \midrule
            \textbf{Other} & & & \\
            \midrule
            \cite{biffi2018learning} & CVAE & & ACDC, Private & MR   & \checkmark & - \\
            \cite{biffi2018learning} 	& CVAE & & Private & MR & \checkmark & Dice, Hausdorff\\
            \cite{volokitin2020modelling}& Slice-to-3D-VAE & & HCP & MR & \checkmark & MMD, MS-SSIM\\
            \cite{huang2022biomarkers}	& GS-VDAE & & MLSP & MR & & Acc.\\
            \cite{amirrajab2022pathology} & VAE-CGAN & Hybrid (V~+~G) & ACDC & MR & $\checkmark$ & -\\
            \cite{beetz2022combined}	& MM-VAE  & & UK Biobank & MR & $\checkmark$ & MMD\\
            \cite{sundgaard2022multi}	& DM-VAE & & Private & Otoscopy & & -\\
		\bottomrule
		\end{tabularx}
	\end{adjustwidth}
	\noindent{\footnotesize{\textsuperscript{1} OCT        stands for ``esophageal optical coherence             tomography''. V = variational autoencoders, G = generative adversarial networks.}}
\end{table}
 
\subsection{Diffusion Models}
In their study, Pinaya et al. \cite{pinaya2022brain} introduce a new approach for generating high-resolution 3D MR images using a latent diffusion model (LDM) \cite{rombach2022high}. LDMs are a type of generative model that combine autoencoders and diffusion models to synthesize new data. The autoencoder component of the LDM compresses the input data into a lower-dimensional latent representation, while the diffusion model component generates new data samples based on this latent representation. The LDM in this work was trained on data from the UK Biobank dataset and conditioned on clinical variables such as age and sex. The authors compare the performance of their LDM to VAE-GAN \cite{vaegan} and LSGAN \cite{mao2016least}, using the Fréchet inception distance \cite{fid} as the evaluation metric. The results show that the LDM outperforms the other models, with an FID of 0.0076 compared to 0.1567 for VAE-GAN and 0.0231 for LSGAN (where a lower FID score indicates a better performance). Even when conditioned on specific variables, the synthetic MRIs generated by this model demonstrate  its ability to produce diverse and realistic brain MRI samples based on the ventricular volume and brain volume. As a valuable contribution to the scientific community, the authors also created a dataset of 100,000 synthetic MRIs that was made openly available for further research.

Fernandez et al. \cite{fernandez2022can} introduce a generative model, named  brainSPADE, for synthesizing labeled brain MRI images that can be used for training segmentation models. The model combines a diffusion model with a VAE-GAN, with the GAN component particularly utilizing SPADE normalization to incorporate the segmentation mask. The model consists of two components: a segmentation map generator and an image generator. The segmentation map generator is a VAE that takes as input a segmentation map, then encodes and builds a latent space from it. To focus on semantic information and disregard insignificant details, the latent code is then diffused and denoised using LDMs. This creates an efficient latent space that emphasizes meaningful information while filtering out noise and other unimportant details. A VAE decoder then generates an artificial segmentation map from this latent space. The image generator is a SPADE model that builds a style latent space from an arbitrary style and combines it with the artificial segmentation map to decode the final output image. The performance of the brainSPADE model is evaluated on a segmentation task using nnU-Net \cite{isensee2018nnu}, and the results show that the model performs comparably when trained on synthetic data compared to when it is trained on real data, and that using a combination of both significantly improves the model's performance.

Lyu and Wang \cite{lyu2022conversion} conducted a study that investigated the use of diffusion models for image translation in medical imaging, specifically the conversion of MRI to CT scans. In their study, the authors utilized two diffusion-based approaches: the conditional DDPM and conditional score-based model which utilizes stochastic differential equations \cite{song2020score}. These methods involved conditioning the reverse process on T2-weighted MRI images. To evaluate the performance of these diffusion models in comparison to other methods (conditional WGAN and U-Net), the authors conducted experiments on the Gold Atlas male pelvis dataset \cite{pelvicdataset} using three novel sampling methods and compared the results to those obtained using GAN- and CNN-based approaches. The results indicated that the diffusion models outperformed both the GAN- and CNN-based methods in terms of structural similarity index (SSIM) and peak signal-to-noise ratio (PNSR).

We present a summary of the relevant studies utilizing diffusion models for medical image augmentation in Table \ref{tab:dm}. This table includes details about the dataset, imaging modality, and evaluation metrics used in each study, as well as the specific diffusion model employed. Upon examining this table, we notice that all the studies included are relatively recent, with the earliest study dating back to 2022. This suggests that diffusion models have gained increasing attention in the field of medical image augmentation and synthesis in recent years. Additionally, we see that in 2022, diffusion models  received more attention for these tasks compared to GANs and VAEs, highlighting their growing popularity and potential for use in various scenarios.

\begin{table}[H]
\caption{Overview of the diffusion-model-based architectures for medical image augmentation that have been published to date (to our knowledge, no such studies were released before 2022). The table includes the reference, architecture name, and hybrid status (if applicable), indicating the combination of VAEs, GANs, and DMs used in each study. The table provides a useful summary of the current state of the art in this area and can help guide researchers in selecting appropriate approaches for their specific needs.\label{tab:dm}} 
\setlength{\cellWidtha}{\fulllength/7-2\tabcolsep-0.2in}
\setlength{\cellWidthb}{\fulllength/7-2\tabcolsep+0.2in}
\setlength{\cellWidthc}{\fulllength/7-2\tabcolsep-0in}
\setlength{\cellWidthd}{\fulllength/7-2\tabcolsep-0in}
\setlength{\cellWidthe}{\fulllength/7-2\tabcolsep-0in}
\setlength{\cellWidthf}{\fulllength/7-2\tabcolsep-0.5in}
\setlength{\cellWidthg}{\fulllength/7-2\tabcolsep+0.5in}
	   \begin{adjustwidth}{-\extralength}{0cm}
		\begin{tabularx}{\fulllength}{>{\raggedright\arraybackslash}m{\cellWidtha}>{\raggedright\arraybackslash}m{\cellWidthb}>{\raggedright\arraybackslash}m{\cellWidthc}>{\raggedright\arraybackslash}m{\cellWidthd}>{\raggedright\arraybackslash}m{\cellWidthe}>{\raggedright\arraybackslash}m{\cellWidthf}>{\raggedright\arraybackslash}m{\cellWidthg}}
        
		\toprule 
%
            \textbf{Reference}	& 
            \textbf{Architecture} &
            \textbf{Hybrid Status} & 
            \textbf{Dataset} &
            \textbf{Modality} &
            \textbf{3D} &
            \textbf{Eval. Metrics}\\
		\midrule 
            \textbf{Classification} & & & & & & \\
            \midrule
		\cite{pinaya2022brain} &     
            CLDM & & UK Biobank & MR & \checkmark & FID, MS-SSIM\\
            \cite{dorjsembe2022three}  & DDPM & & ICTS & MR & \checkmark & MS-SSIM\\
            \cite{packhauser2022generation} 	& LDM &  & CXR8 & X-ray & & AUC\\
            \cite{moghadam2022morphology}	& MF-DPM & & TCGA & Dermoscopy & & Recall\\
            \cite{chambon2022roentgen}	& RoentGen & Hybrid (D~+~V) & MIMIC-CXR & X-ray & & Accuracy\\
            \cite{wolleb2022swiss}	& IITM-Diffusion & & BraTS2020 & MR & & - \\
            \cite{sagers2022improving}	& DALL-E2 & & Fitzpatrick & Dermoscopy & & Accuracy\\
            \cite{peng2022generating}	& CDDPM & & ADNI & MR & \checkmark & MMD, MS-SSIM, FID\\
            \cite{ali2023spot} & DALL-E2 & & Private & X-ray & & - \\
            \cite{saeed2023bi} & DDPM & & OPMR & MR & \checkmark & Acc., Dice\\
            \cite{weber2023cascaded} & LDM & & MaCheX & X-ray & & MSE, PSNR, SSIM\\
 	          \midrule
            \textbf{Segmentation} & & & & & \\
            \midrule
            \cite{khader2022medical}	& DDPM & & ADNI, MRNet, & MR, CT & & Dice\\
            &&& LIDC-IDRI &&&\\ 
            \cite{fernandez2022can}		& brainSPADE & Hybrid (V~+~G~+~D) & SABRE, BraTS2015 & MR & & Dice, Accuracy \\
            &&& OASIS, ABIDE &&& Precision, Recall \\
            \cite{wolleb2022swiss}		& IITM-Diffusion & & BraTS2020 & MR & & - \\

            \midrule
            \textbf{Cross-modal translation} & & & & & \\
            \midrule
            \cite{ozbey2022unsupervised} & SynDiff & Hybrid (D~+~G) & IXI, BraTS2015 & CT $\rightarrow$ MR & & PSNR, SSIM \\
            &&& MRI-CT-PTGA &&& \\
            \cite{meng2022novel}	& UMM-CSGM & & BraTS2019 & FLAIR $\leftrightarrow$ T1 $\leftrightarrow$ T1c $\leftrightarrow$ T2 & & PSNR, SSIM, MAE\\
            \cite{lyu2022conversion}& CDDPM & & MRI-CT-PTGA & CT $\leftrightarrow$ MR &  & PSNR, SSIM \\
            \midrule
            \textbf{Other} & & & & & \\
            \midrule
            \cite{kim2022diffusion} & DDM & & ACDC & MR   & \checkmark & PSNR, NMSE, DICE\\
		\bottomrule
		\end{tabularx}
	\end{adjustwidth}
	\noindent{\footnotesize{Note: V = variational autoencoders, G = generative adversarial networks, D = diffusion models.}}
\end{table}

\section{Key Findings and Implications}\label{sec4}
In this review, we focused on generative deep models applied to medical data augmentation, specifically VAEs, GANs, and diffusion models. These approaches each have their own strengths and limitations, as described by the generative learning trilemma \cite{xiao2021tackling}, which states that it is generally difficult to achieve high-quality sampling, fast sampling, and mode coverage simultaneously. As illustrated in Figure \ref{fig:stats}a, the number of publications on data augmentation using VAEs increases by approximately 81\% from 2017 to 2022, while the number using GANs has remained relatively stagnant. This trend may be due to the fact that most possible fields of research using GANs have already been explored, making it difficult to go beyond current methods using these architectures. However, we have also seen an increase in the use of more complex architectures combining multiple generative models \cite{ahmad2022brain,yang2021synthesizing}, which have shown promising results in terms of both quality and mode coverage. On the other hand, the number of studies using diffusion models has drastically increased starting from 2022, and these models have shown particular potential for synthesizing high-quality images with good mode coverage \cite{kazerouni2022diffusion}. 

{\color{black} Basic data augmentation operators such as Gaussian noise addition, cropping, and padding are commonly used to augment data and generate new images for training~ \cite{krizhevsky2017imagenet}. However, the complex structures of medical images, which encompass anatomical variation and irregular tumor shapes, may render these basic operations unsuitable, resulting in the production of irrelevant images that disrupt the logical image structure \cite{chlap2021review}, and additionally, can lead to image deformations and the generation of aberrant data that can adversely impact the model performance. One basic data augmentation operator that is not well suited for medical images is flipping images, which can sometimes cause anatomical inconsistencies \cite{abdollahi2020data}. To overcome this issue, deformable augmentation techniques have been introduced, such as random displacement fields and spline interpolation, to augment the data in a more realistic way. These techniques have proved to be useful \cite{chlap2021review}; however, they are strongly dependent on the data and limited in some cases. Recent advances in deep learning have led to the development of generative models that can be trained to generate realistic images and simulate the underlying data distribution. These synthesized images are more truthful than those generated using traditional data augmentation techniques. They guarantee a better coherence of the general structure of medical images and greater variability, providing a more effective way to generate realistic and diverse data.}

{\color{black} The use of GANs in medical imaging, as seen in Table \ref{tab:gan}, has been widespread and applied to a variety of modalities and datasets, demonstrating their versatility and potential for various applications within the field. When it comes to classification, DCGAN and WGAN have been the most-commonly used architectures and are considered safe bets in this domain. For example, Zhuang et al. \cite{zhuang2019fmri} demonstrated a 3\% accuracy improvement in generating fMRIs using an improved WGAN. These architectures, with their capacity for high-quality generation and good mode coverage, offer significant potential for the generation of synthetic images for medical imaging classification. In the case of segmentation and translation, the architectures that have shown the most promise include Pix2Pix, CycleGAN, and SPADE, all of which have proven their potential for conditional generation and cross-modal translation. Platscher et al. \cite{platscher2020image} conducted a comparative study of these three architectures, demonstrating their capacity to generate high-quality images suitable for medical image segmentation and translation tasks (improvement of 9.1\% in Dice score). These architectures can significantly reduce the need for manual annotation of medical images and thus significantly reduce the time and cost required for data annotation.}

{\color{black} On the other hand, VAEs have been utilized in fewer studies for medical image augmentation, as shown in Table \ref{tab:vae}. They have been employed in other tasks such as reconstruction, as demonstrated by Biffi et al. \cite{biffi2018learning} and Volokitin et al. \cite{volokitin2020modelling}, who used CVAE for 3D volume reconstruction, and interpretability of features, as exemplified by Hyang et al. \cite{huang2022biomarkers}, who identified biomarkers using VAEs. Furthermore, VAEs are often used in hybrid architectures with adversarial learning techniques. The most promising architectures include PAVAE \cite{huo2022brain} and IntroVAE \cite{huang2018introvae}, alongside conditional VAEs, for various purposes including classification, segmentation, and translation tasks. However, while VAEs have shown potential in these areas, there is still room for improvement. One study that particularly shows promising results is that of Chadebec and Allassonnière~\cite{chadebec2021data}, who propose to model the latent space of a VAE as a Riemannian manifold, allowing high-quality image generation comparable to GANs. Chadebec and Allassonnière \cite{chadebec2021data} demonstrated an improvement of 8\% in accuracy using synthetic images generated with their proposed VAE model. Nevertheless, this architecture requires a high computational cost and time, which is a significant drawback in practical applications.}

\begin{figure}[H]
\includegraphics[width=13cm]{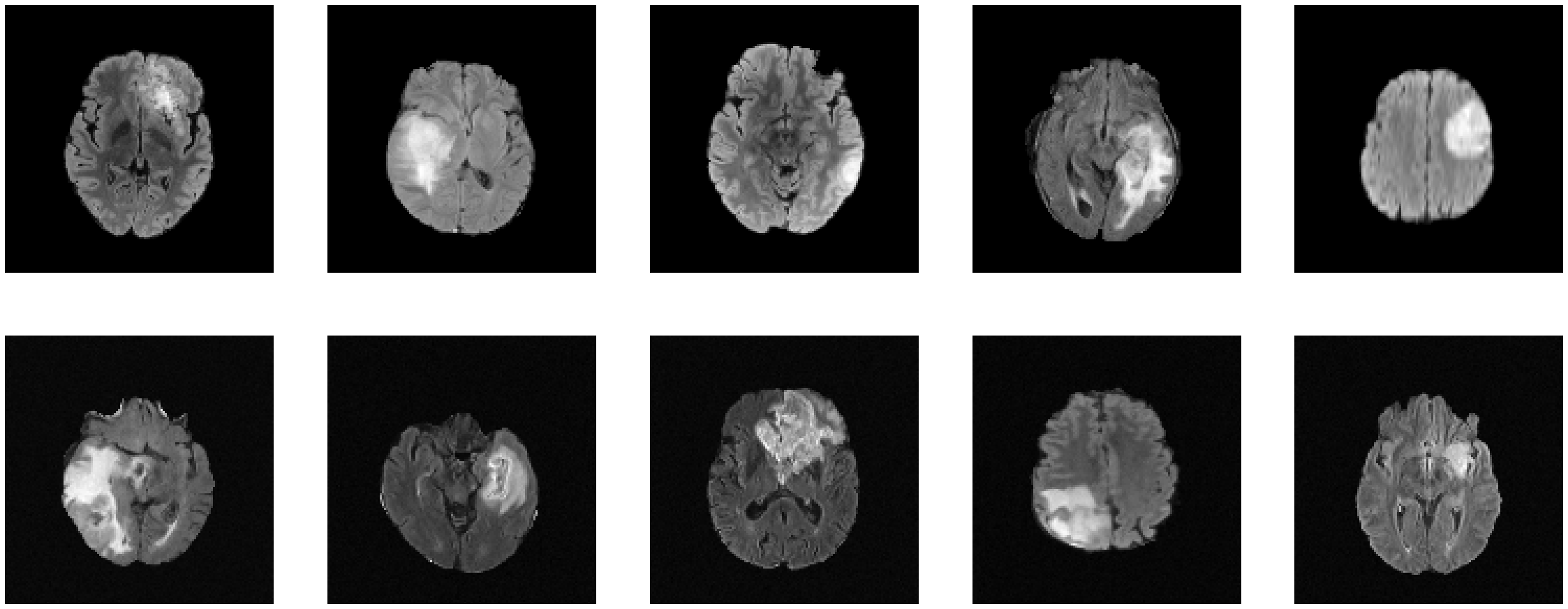}
\caption{{\color{black} Synthesized MRIs using a diffusion-based probabilistic model (DDPM) \cite{ho2020denoising} trained on the BraTS2020 dataset. The first row shows a sample of original images, while the second row shows a sample of synthesized images generated using the DDPM.} \label{fig:dm}}
\end{figure}

Table \ref{tab:dm} presents a summary of the relevant studies utilizing diffusion models for medical image augmentation. These studies, all of which are relatively recent, with the earliest dating back to 2022, suggest that diffusion models have gained increasing attention in medical image augmentation and synthesis in recent years. Furthermore, in 2022, diffusion models have been the most-commonly used generative models for medical image augmentation compared to GANs and VAEs, highlighting their growing popularity and potential for use in various scenarios. {\color{black} Of the diffusion models studied, DDPM and LDM are the most prevalent, alongside conditional variants such as CDDPM \cite{lyu2022conversion} and CLDM \cite{pinaya2022brain}. Notably, the difference between LDM and DDPM is the ability of LDM to model long-range dependencies within the data by constructing a low-dimensional latent representation and diffusing it, while DDPMs apply the diffusion process directly to the input images. This can be especially useful for medical image augmentation tasks that require capturing complex patterns and structures. For instance, Saeed et al. \cite{saeed2023bi} demonstrated the capacity of LDM conditioned on text for a task of lesion identification, achieving an accuracy improvement of 5.8\%. These findings suggest that diffusion models have a promising potential for future medical image augmentation and synthesis research. To further exemplify the potential of diffusion models in generating realistic medical images, we present in Figure \ref{fig:dm} a set of synthesized MRI images using a DDPM. These generated images exhibit high visual fidelity and are almost indistinguishable from the real images. One of the reasons for this high quality is  the DDPM's ability to model the diffusion process of the image density function. By doing so, the DDPM can generate images with increased sharpness and fine details, as seen in the synthesized MRI images.}

These studies have covered a range of modalities, including MRI, CT, and ultrasound, as well as dermoscopy and otoscopy. Classification is the most common downstream task targeted in these studies, but there have also been multiple state-of-the-art solutions proposed for more complex tasks such as generating multimodal missing images (e.g., from CT to MRI) and multi-contrast MRI images. In order to provide ground truth segmentation masks for tasks such as segmentation, most studies have explored the field of conditional synthesis. This allows for greater control over the synthesized images and can help to stabilize training \cite{mirza2014conditional}, as the model is given explicit guidance on the desired output. For our discussion on medical image augmentation, we have also compiled two summary tables to provide a comprehensive overview of the datasets and metrics used in the reviewed studies. Table \ref{tab:datasets} presents a summary of the datasets used in the reviewed studies. This table includes information about the title of the dataset, a reference, and a link to the public repository if available, as well as the studied modality and anatomy. From examining this table, we see that MRI is the most-commonly used modality, followed by CT. In terms of anatomy, brain studies dominate, with lung studies coming in second. It is worth noting that the BraTS dataset is widely used across multiple studies, highlighting its importance in the field. Additionally, we notice the presence of private datasets in this table, which is not surprising given that many medical studies are associated with specific medical centers and may not be publicly available. When we consider the state of the art of medical imaging studies (see Figure \ref{fig:stats}b), we notice that the PET and ultrasound modalities are less represented compared to the others. One reason for the scarcity of PET studies is the limited availability of nuclear doctors compared to radiologists. Nuclear doctors specialize in nuclear medicine, and PET is one such imaging modality that uses radioactive tracers to produce 3D images of the body. Due to the limited number of nuclear doctors, there are fewer medical exams that use PET, leading to less publicly available data for research purposes \cite{amyar2020radiogan}. On the other hand, ultrasound is an operator-dependent modality and requires a certain level of field knowledge. Additionally, ultrasound is not as effective as other modalities such as CT and MRI in detecting certain pathologies, which may also contribute to its lower representation in the state of the art. Despite these limitations, both PET and ultrasound remain important imaging modalities in clinical practice, and future research should aim to explore their full potential in the field of medical imaging.

\begin{table}[H]
\tablesize{\small}
\caption{Summary of the datasets utilized in various publications of deep generative models, organized by modality and body part. For each dataset, the corresponding availability is indicated as public, private, or under certain conditions (UC). Additionally, if a public link for the dataset is available, it is provided.\label{tab:datasets}} 
\setlength{\cellWidtha}{\fulllength/6-2\tabcolsep-0.1in}
\setlength{\cellWidthb}{\fulllength/6-2\tabcolsep-0.4in}
\setlength{\cellWidthc}{\fulllength/6-2\tabcolsep-0.3in}
\setlength{\cellWidthd}{\fulllength/6-2\tabcolsep+1.1in}
\setlength{\cellWidthe}{\fulllength/6-2\tabcolsep-0.2in}
\setlength{\cellWidthf}{\fulllength/6-2\tabcolsep-0.1in}
	   \begin{adjustwidth}{-\extralength}{0cm}
		\begin{tabularx}{\fulllength}{>{\raggedright\arraybackslash}m{\cellWidtha}>{\raggedright\arraybackslash}m{\cellWidthb}>{\raggedright\arraybackslash}m{\cellWidthc}>{\raggedright\arraybackslash}m{\cellWidthd}>{\raggedright\arraybackslash}m{\cellWidthe}>{\raggedright\arraybackslash}m{\cellWidthf}
		}
        
		\toprule 
            \textbf{Abbreviation} & \textbf{Reference}	& \textbf{Availability} & \textbf{Dataset} & \textbf{Modality} & \textbf{Anatomy}\\
		\midrule
            {ADNI} & & UC & Alzheimers disease \mbox{neuroimaging Initiative} & MR, PET & Brain\\
            {BraTS2015} & & Public & Brain tumor segmentation challenge & MR & Brain\\
            {BraTS2016} & & Public & Brain tumor segmentation challenge & MR & Brain\\
            {BraTS2017} & & Public & Brain tumor segmentation challenge & MR & Brain\\
            {BraTS2019} & & Public & Brain tumor segmentation challenge & MR & Brain\\
            {BraTS2020} & & Public & Brain tumor segmentation challenge & MR & Brain\\
            {IEEE CCX} & & Public & IEEE Covid Chest X-ray dataset & X-ray & Lung\\
            {UK Biobank} & & UC & UK Biobank & MR & Brain, Heart\\
            {NIHPCT} & & Public &National Institutes of Health Pancreas-CT dataset & CT & Kidney\\
            {DataDecathlon} & & Public & Medical Segmentation \mbox{Decathlon dataset} & CT & Liver, Spleen\\
            {MIDAS} & \cite{midasdataset} & Public & Michigan institute for data science & MR & Brain\\
            {IXI} & & Public & Information e\textbf{X}traction from \mbox{Images Dataset} & MR & Brain\\
            {DRIVE} &\cite{drivedataset} &  Public & Digital Retinal Images for \mbox{Vessel Extraction} & Fundus photography & Retinal fundus\\
            {ACDC}& \cite{acdcdataset} &  Public & Automated Cardiac \mbox{Diagnosis Challenge} & MR & Heart\\
            {MRI-CT PTGA} & \cite{pelvicdataset} &  Public & MRI-CT Part of the Gold Atlas project & CT, MR & Pelvis\\
            {ICTS} & \cite{kwon2019generation} & Public & National Taiwan University Hospital’s Intracranial Tumor \mbox{Segmentation dataset} & MR & Brain\\
            {CXR8} & \cite{wang2017chestx} & Public & ChestX-ray8 & X-ray & Lung\\
            {C19CT} & & Public & COVID-19 CT segmentation dataset & CT & Lung\\
            {TCGA} & & Private & The Cancer Genome Atlas Program
             & Microscopy & -\\
             {UKDHP} & \cite{bai2015bi} & UC & UK Digital Heart Project & MR & Heart\\
             {SCR} & \cite{van2006segmentation} & Public & SCR database : Segmentation in \mbox{Chest Radiographs} & X-ray & Lung\\
             {HCP} & \cite{van2013wu} & Public & Human connectom project dataset & MR & Brain\\
            {AIBL} &
            & UC & Australian Imaging Biomarkers and Lifestyle Study of Ageing & MR, PET & Brain\\
            {OpenfMRI} & & Public & OpenfMRI & MR & Brain\\
            {IBC} & & Public & Individual Brain Charting & MR & Brain\\
{NeuroSpin} & & Private & Institut des sciences du vivant \mbox{Frédéric Joliot} & MR & Brain\\

          {OASIS} & & Public & The Open Access Series of \mbox{Imaging Studies} & MR & Brain\\
            {APCXR} & \cite{kermany2018identifying} & Public & The anterior-posterior Chest \mbox{X-Ray dataset} & X-ray & Lung\\
            {Fitzpatrick} & \cite{groh2021evaluating} & Public & Fitzpatrick17k dataset & Dermoscopy & Skin\\
            
                        {ISIC} & & Public & The International Skin Imaging Collaboration dataset & Dermoscopy & Skin\\

\bottomrule
		\end{tabularx}
	\end{adjustwidth}
\end{table}
 
\begin{table}[H]\ContinuedFloat
\tablesize{\small}
\caption{\textit{Cont.} \label{tab:datasets}} 
\setlength{\cellWidtha}{\fulllength/6-2\tabcolsep-0.1in}
\setlength{\cellWidthb}{\fulllength/6-2\tabcolsep-0.4in}
\setlength{\cellWidthc}{\fulllength/6-2\tabcolsep-0.3in}
\setlength{\cellWidthd}{\fulllength/6-2\tabcolsep+1.1in}
\setlength{\cellWidthe}{\fulllength/6-2\tabcolsep-0.2in}
\setlength{\cellWidthf}{\fulllength/6-2\tabcolsep-0.1in}
	   \begin{adjustwidth}{-\extralength}{0cm}
		\begin{tabularx}{\fulllength}{>{\raggedright\arraybackslash}m{\cellWidtha}>{\raggedright\arraybackslash}m{\cellWidthb}>{\raggedright\arraybackslash}m{\cellWidthc}>{\raggedright\arraybackslash}m{\cellWidthd}>{\raggedright\arraybackslash}m{\cellWidthe}>{\raggedright\arraybackslash}m{\cellWidthf}
		}
        
		\toprule 
            \textbf{Abbreviation} & \textbf{Reference}	& \textbf{Availability} & \textbf{Dataset} & \textbf{Modality} & \textbf{Anatomy}\\
		\midrule           
  

            {DDSM}& & Public & The Digital Database for Screening Mammography & CT & Breast\\
            {CBIS-DDMS}& & Public & Curated Breast Imaging Subset \mbox{of DDSM} & CT & Breast\\
            {LIDC-IDRI} & & Public & The Lung Image Database Consortium (LIDC) and Image Database Resource Initiative (IDRI) & CT & Lung\\
            {COVID-CT} & \cite{yang2020covid} & Public & - & CT & Lung\\
            {SARS-COV2} & \cite{soares2020sars} & Public &  & CT & Lung\\
            {MIMIC-CXR} & \cite{johnson2019mimic} & Public & Massachusetts Institute of Technology & CT & Lung\\

            {PPMI} & & Public & Parkinson's Progression \mbox{Markers Initiative} & MR & Brain\\

            {ADHD}& & Public & Attention Deficit \mbox{Hyperactivity Disorder} & MR & Brain\\
            {MRNet}& & Public & MRNet dataset & MR & Knee\\
            {MLSP}& & Public & MLSP 2014 Schizophrenia Classification Challenge & MR & Brain\\
            {SABRE} & \cite{jones2020cohort} & Public & The Southall and Brent
            Revisited cohort & MR & Brain, Heart\\
            {ABIDE} &  & Public & The Autism Brain Imaging \mbox{Data Exchange} & MR & Brain\\
            {OPMR} & \cite{saha2022artificial} & Public & Open-source prostate MR data & MR & Pelvis\\
            {MaCheX} & \cite{weber2023cascaded} & Public & Massive Chest X-ray Dataset & X-ray & Lung\\
		\bottomrule
		\end{tabularx}
	\end{adjustwidth}
\end{table}

\begin{figure}[H]
\begin{adjustwidth}{-\extralength}{0cm}
\centering
\includegraphics[width=17cm]{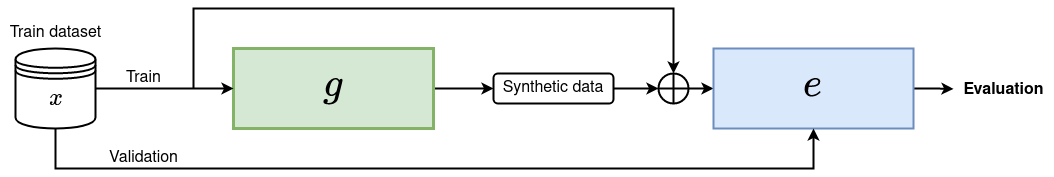}
\end{adjustwidth}
\caption{Illustration of the augmentation pipeline for a generative-model-based data augmentation. The input data, $x$, are fed into the generative model, $g$, which synthesizes additional data samples to augment the training set. The downstream architecture, $e$, which may take the form of a convolutional neural network or U-Net, is then trained on a combination of the synthesized data and real data from the training set. The training set is split into training and validation sets, where the validation set contains only real data for evaluation purposes. After training, the model can be evaluated using various test sets.} \label{fig:aug_flow}
\end{figure}

Second, Table \ref{tab:metrics} provides a summary of the metrics used to evaluate the performance of the various models discussed in the review. It is clear from this table that a variety of metrics are employed, ranging from traditional evaluation measures to more recent ones. Currently, many studies rely on shallow metrics such as the mean absolute error, peak signal-to-noise ratio \cite{kynkaanniemi2019improved}, or structural similarity \cite{ssim}, which do not accurately reflect the visual quality of the image. For instance, while optimizing pixel-wise loss can produce a clearer image, it may result in lower numerical scores compared to using adversarial loss \cite{ledig2017photo}. To address this challenge, researchers have proposed different methods for evaluation. The most well-known approach is to validate the quality of the generated samples through downstream tasks such as segmentation or classification. {\color{black} An overview of the augmentation process using a downstream task is depicted in Figure \ref{fig:aug_flow}}. Another approach is to use deep-learning-based metrics such as the learned perceptual image patch similarity (LPIPS) \cite{zhang2018unreasonable}, Fréchet inception distance (FID) \cite
{fid}, or inception score (IS) \cite{salimans2016improved}, which are designed to better reflect human judgments of image quality. These deep-learning-based metrics take into account not only pixel-wise similarities, but also high-level features and semantic information in the images, making them more effective in evaluating the visual quality of the generated images. LPIPS, for instance, measures the perceptual similarity between two images by using a pretrained deep neural network. FID and IS are other popular deep-learning-based metrics for image generation, and they have been widely used in various image generation tasks to assess the quality and diversity of the generated samples. However, these metrics may not always align perfectly with human perception, and further studies are needed to assess their effectiveness for different types of medical images.

\begin{table}[H]
\tablesize{\footnotesize}
\caption{Summary of quantitative measures used in the reviewed publications.\label{tab:metrics}} 
\setlength{\cellWidtha}{\fulllength/4-2\tabcolsep-1in}
\setlength{\cellWidthb}{\fulllength/4-2\tabcolsep-1in}
\setlength{\cellWidthc}{\fulllength/4-2\tabcolsep-0.5in}
\setlength{\cellWidthd}{\fulllength/4-2\tabcolsep+2.5in}
	   \begin{adjustwidth}{-\extralength}{0cm}
		\begin{tabularx}{\fulllength}{>{\raggedright\arraybackslash}m{\cellWidtha}>{\raggedright\arraybackslash}m{\cellWidthb}>{\raggedright\arraybackslash}m{\cellWidthc}>{\raggedright\arraybackslash}m{\cellWidthd}
		}
        
		\toprule 

            \textbf{{\footnotesize Abbrv.}}	& 
            \textbf{{\footnotesize Reference}} &
            \textbf{{\footnotesize Metric Name}} & 
            \textbf{{\footnotesize Description}} \\
		\midrule 
        Dice & \cite{sudre2017generalised} & Sørensen–Dice coefficient & A measure of the similarity between two sets of data, calculated as twice the size of the intersection of the two sets divided by the sum of the sizes of the two sets\\
        Hausdorff & \cite{rockafellar2009variational} & Hausdorff distance & A measure of the similarity between two sets of points in a metric space\\
        FID & \cite{fid} & Fréchet inception distance & A measure of the distance between the distributions of features extracted from real and generated images, based on the activation patterns of a pretrained inception model\\
        IS & \cite{salimans2016improved} & Inception score & A measure of the quality and diversity of generated images, based on the activation patterns of a pretrained Inception model\\
        MMD & \cite{gretton2012kernel} & Maximum mean discrepancy & A measure of the difference between two probability distributions, defined as the maximum value of the difference between the two means\\
        1-NN & \cite{cover1967nearest} & 1-nearest neighbor score & A method for classification or regression that involves finding the data point in a dataset that is most similar to a given query point\\
        (MS-)SSIM & \cite{ssim}& (Multi-scale) structural similarity & A measure of the similarity between two images based on their structural information, taking into account luminance, contrast, and structure.\\
        MS & \cite{bounliphone2015test} & Mode score & A measure of the quality of samples generated with two probabilistic generative models based on the difference in maximum mean discrepancies between a reference distribution and simulated distribution\\
        WD & \cite{vaserstein1969markov} & Wasserstein distance & A measure of the distance between two probability distributions, defined as the minimum amount of work required to transform one distribution into the other\\
        PSNR & \cite{kynkaanniemi2019improved} & Peak signal-to-noise ratio & A measure of the quality of an image or video, based on the ratio between the maximum possible power of a signal and the power of the noise that distorts the signal\\
        (N)MSE & - & (Normalized) mean squared error & A measure of the average squared difference between the predicted and \mbox{actual values}\\
        Jacc. & \cite{sudre2017generalised} & Jaccard index & A measure of the overlap between two sets of data, calculated as the ratio of the area of intersection to the area of union\\
        MAE & - & Mean absolute error & A measure of the average magnitude of the errors between the predicted and \mbox{actual values}\\
        AUC & \cite{fawcett2006introduction} & Area under the curve & A measure of the performance of a binary classifier, calculated as the area under the receiver operating characteristic curve\\
        LPIPS & \cite{zhang2018unreasonable} & Learned perceptual image patch similarity & An evaluation metric that measures the distance between two images in a perceptual space based on the activation of a deep CNN\\
        KLD & \cite{nguyen2010estimating} & Kullback--Leibler divergence & A measure of the difference between two probability distributions, often used to compare the similarity of the distributions, with a smaller KL divergence indicating a greater similarity\\
        VIF & \cite{vif} & Visual information fidelity &  A measure that quantifies the Shannon information that is shared between the reference and the distorted image\\
        UQI & \cite{uqi} & Universal quality index & A measure of the quality of restored images. It is based on the principle that the quality of an image can be quantified using the correlation between the original and restored images\\
		\bottomrule
		\end{tabularx}
	\end{adjustwidth}
\end{table}
 \vspace{-12pt}

Despite the advancements made by generative models in medical data augmentation, several challenges still remain. A common issue in GANs, known as mode collapse, occurs when the generator only produces a limited range of outputs, rather than the full range of possibilities. While techniques such as minibatch discrimination and the incorporation of auxiliary tasks \cite{salimans2016improved} have been suggested as potential solutions, further research is needed to effectively address this issue. In addition, there is a balance to be struck between the sample quality and the generation speed, which affects all generative models. GANs are known for their ability to generate high-quality samples quickly, allowing them to be widely used in medical imaging and data augmentation \cite{yi2019generative,tavse2022systematic,ali2022role}. Another approach for stabilizing the training of GANs is to use WGAN \cite{arjovsky2017wasserstein}. WGAN improves upon the original GAN by using the Wasserstein distance instead of the Jensen--Shannon divergence as the cost function for training the discriminator network. While these approaches have demonstrated success in improving GAN images and partially addressing mode collapse and training instability, there is still room for improvement. Diffusion models have overshadowed GANs during the latest years, particularly due to the success of text-to-image generation architectures such as DALL-E \cite{ramesh2021zero}, Imagen \cite{saharia2022photorealistic}, and stable diffusion \cite{rombach2022high}. These diffusion models naturally produce more realistic images than GANs. However, in our view, GANs have only been set aside and not entirely disregarded. With the recent release of GigaGAN~ \cite{kang2023scaling} and StyleGAN-T \cite{sauer2023stylegan}, GANs have made a resurgence by producing comparable or even better results than diffusion models. This renewed interest in GANs demonstrates the continued relevance of this approach to image generation and indicates that GANs may still have much to offer in advancing the field. Future research could explore hybrid models that combine the strengths of both GANs and diffusion models to create even more realistic and high-quality images.

\begin{figure}[H]
\includegraphics[width=12cm]{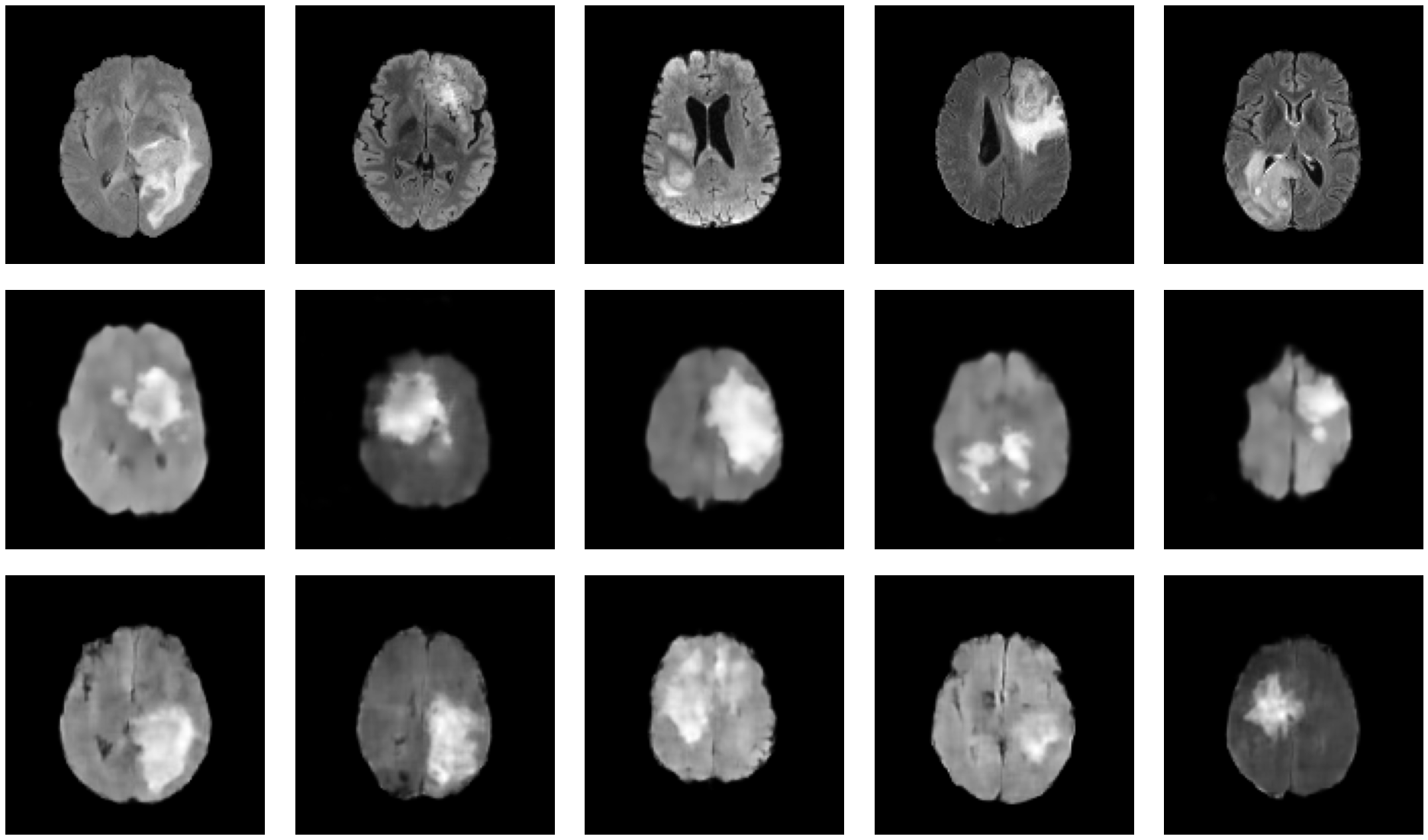}
\caption{Figure presenting a comparison between synthesized MRIs generated by a VAE and a Hamiltonian VAE \cite{caterini2018hamiltonian}. Both models were trained on a limited training set of 100 images from BraTS2020 Challenge dataset. The first row showcases original images, while the second and third rows present synthesized images generated by the VAE and Hamiltonian VAE, respectively. While the images generated by both models appear slightly fuzzy, the Hamiltonian VAE demonstrates enhanced performance in generating realistic images. This comparison highlights the robustness of the VAE and Hamiltonian VAE for generating new images from a small dataset \cite{delgado2021deep}.\label{fig:vae}}
\end{figure}

VAEs have not gained as much attention in the medical imaging field, due to their tendency to produce blurry and hazy generated images. However, some studies have used conditional VAEs or hybrid architectures to address this issue and improve the quality of the samples produced. Researchers are therefore exploring the use of hybrid models that combine the strengths of multiple generative models, as well as improved VAE variations that offer enhanced image quality. Hybrid architectures, such as VAE-GANs~\cite{vaegan}, have demonstrated the potential to partially address the issues of both VAEs and GANs, allowing a better-quality generation and good mode coverage. {\color{black} Interestingly, recent research has even combined all three generative models into a single pipeline \cite{fernandez2022can}. This study has shown comparable results on a segmentation task when using a fully synthetic dataset compared to using the real dataset. These promising results suggest that hybrid architectures could open up new possibilities.} However, these models can be complex and challenging to train, and more research is needed to fully realize their potential. In fact, many VAEs used in medical imaging are hybrid architectures, as they offer a good balance between the strengths and weaknesses of both VAEs and GANs \cite{fernandez2022can,terzopoulos2019multi}. {\color{black} It is important to note that VAEs have an advantage over GANs in operating better with smaller datasets due to the presence of an encoder \cite{delgado2021deep}, which can extract relevant features from the input images and significantly reduce the search space required for generating new images through the process of reconstruction. This feature also makes VAEs a form of dimensionality reduction, and the representation obtained by the encoder can provide a better starting point for the decoder to approximate the real data distribution more accurately. In contrast, GANs have a wider search space, which may lead to challenges in learning features effectively. For instance, we show in Figure \ref{fig:vae} a comparison between synthesized MRIs using vanilla VAE \cite{kingma2013auto} and the Hamiltonian VAE \cite{caterini2018hamiltonian}. In addition to the advantage of operating better with smaller datasets, VAEs also offer a disentangled, interpretable, and editable latent space. This means that the encoded representation of an input image can be separated into independent and interpretable features, allowing for better understanding and manipulation of the underlying data.} Another option is the use of improved variants of VAEs, which have been proposed to generate high-quality images. There has been limited exploration of improved VAE variants such as VQ-VAE2 \cite{vqvae2}, IAF-VAE \cite{iafvae}, or Hamiltonian VAE \cite{caterini2018hamiltonian} in the medical imaging field, but these variants have shown promise in generating high-quality images in other domains. It may be worth exploring their potential for medical image augmentation, as they offer the possibility of improving the quality of the generated images without sacrificing other important characteristics such as fast sampling and good mode coverage.

Diffusion models have more recently been applied to medical imaging \cite{kazerouni2022diffusion}, and some studies have demonstrated high-quality results \cite{dorjsembe2022three}. These models are capable of synthesizing highly realistic images and have a good mode coverage while keeping the training stable, but suffer from a long sampling time due to the high number of steps in the diffusion process. This limitation may be less significant in medical imaging applications, which are not typically used in real time, but researchers are likely to continue working on optimizing diffusion models for faster sampling. It may also be possible to trade off some sample quality for faster sampling in diffusion models \cite{xiao2021tackling}, as realism is a key requirement for data augmentation in medical imaging. For example, Song et al. \cite{song2020denoising} proposes a variant of diffusion models called Training-free Denoising Diffusion Implicit Model (DDIM) aimed to speed up the sampling process by replacing the Markovian process with a non-Markovian one in the DDPM. This resulted in a faster sampling procedure that did not significantly compromise the quality of the samples. Fast Diffusion Probabilistic Model (FastDPM) \cite{kong2021fast} introduces the concept of a continuous diffusion process with smaller time steps in order to reduce the sampling time. These efforts to improve the efficiency of diffusion models demonstrate the ongoing interest in finding ways to balance the sample quality and generation speed in medical imaging applications.

\textls[-25]{There are several other factors to consider when discussing the use of generative models for medical data augmentation. One important factor is the incorporation of domain-specific techniques and knowledge into the design of these models \cite{he2022effect}. By incorporating knowledge of anatomy and physiology, for example, researchers can improve the realism and utility of the generated data. Another important factor is the ethical considerations of using synthetic data for medical applications, including the potential for biased or unrealistic generated data and the need for proper validation and testing. To further improve the performance and efficiency of medical data augmentation, researchers are also exploring the use of generative models in combination with other techniques such as transfer learning \cite{talo2019application,shorten2019survey} or active learning \cite{ren2021survey,rahimi2021addressing}. The role of interpretability and explainability in these models is also important to consider, particularly in the context of clinical decision making and regulatory requirements. In addition to data augmentation, generative models have the potential to be used for other medical applications such as generating synthetic patient records or synthesizing medical images from non-image data~ \cite{chambon2022roentgen}.}


\section{Conclusions}\label{sec5}
{\textls[-25]{In this review, we examine the use of deep generative models for medical image augmentation. The limited availability of training data remains a major challenge in medical image analysis with deep learning approaches, which can be addressed by data augmentation techniques. However, traditional techniques still produce limited and unconvincing results. We focus on three types of deep generative models for medical image augmentation, VAEs, GANs, and DMs, and provide an overview of the current state of the art in each of these models. While deep generative models offer several advantages over traditional data augmentation techniques, including the ability to generate realistic new images that capture the underlying distribution of the training dataset, they also have some limitations. VAEs offer the ability to learn a meaningful and disentangled representation of the data, which can be useful for interpretability and latent space addition. 
 Despite these advantages, VAEs may produce fuzzy images that lack important details, which can be especially problematic in medical imaging. To address this limitation, improved VAE variants have been developed, such as vector quantized VAE, which uses powerful priors to generate synthetic samples with higher coherence and fidelity. Another approach involves combining VAEs with adversarial learning to improve the level of detail in the generated images. Alternatively, GANs have been found to generate high-quality images with fine details, and can be memory-efficient due to their upsampling-only architecture. \textls[-15]{However, GANs can be difficult to train and may suffer from mode collapse. Techniques such as WGAN and minibatch discrimination can help stabilize GAN training, and increasing the size of the training set can also be effective. Diffusion models have also been shown to generate high-quality images with increased sharpness and fine details, better than previous generative models, but they require significant computational resources to train and may be less interpretable. Researchers are currently exploring ways to reduce the sampling time of diffusion models, such as with progressive distillation, FastDPM, and DDIM variants. Overall, while each approach has its own strengths and weaknesses, continued research and development will be crucial for improving the effectiveness of deep generative models in various applications, including medical imaging. This evaluation of the strengths and limitations of each model can suggest directions for future research in this field including the exploration of hybrid architectures and improved variants, the incorporation of domain-specific knowledge, and the combination with other techniques such as transfer learning or active learning. The aim of this review is to emphasize the potential of deep generative models in enhancing the performance of deep learning algorithms for medical image analysis. By identifying the challenges of the current methods, we seek to increase awareness of the need for further contributions in this field.}}}

\vspace{6pt}

\authorcontributions{Conceptualization, A.K., S.R. and J.L.-L.; methodology, A.K., S.R. and J.L.-L.; software, A.K.; validation, S.R. and J.L.-L.;  writing—original draft preparation, A.K.; writing—review and editing, S.R. and J.L.-L.; supervision, S.R. and J.L.-L. All authors have read and agreed to the published version of the manuscript.
}

\funding{This research received no external funding.
}

\conflictsofinterest{The authors declare no conflict of interest. The funders had no role in the design of the study; in the collection, analyses, or interpretation of data; in the writing of the manuscript; or in the decision to publish the~results.
} 



\begin{adjustwidth}{-\extralength}{0cm}
\reftitle{References}

%

%


%
%
%
\PublishersNote{}
\end{adjustwidth}
\end{document}